\documentclass[12pt,a4paper]{amsart}
\usepackage[all]{xy} 

\usepackage{amssymb} 
\usepackage{eucal}
\usepackage{hyperref}
\numberwithin{equation}{section}

\newtheorem{definition}{Definition}[section]
\newtheorem{lemma}[definition]{Lemma}
\newtheorem{theorem}[definition]{Theorem}
\newtheorem{proposition}[definition]{Proposition}
\newtheorem{corollary}[definition]{Corollary}
\newtheorem{remarkth}[definition]{Remark}

\newenvironment{remark}{\begin{remarkth}\upshape}{\hfill$\diamond$\end{remarkth}}

\renewcommand{\emph}[1]{{\bfseries\itshape{#1}}}

\newcommand{\R}{\mathbb{R}} 

\newcommand{\lcf}{\lbrack\! \lbrack}
\newcommand{\rcf}{\rbrack\! \rbrack}

\renewcommand{\d}{\mathrm{d}^\circ}

\makeatletter
\newcommand\prol{\@ifstar{\@proldf}{\@prolpf}} 
\def\@prolpf{\@ifnextchar[{\@prolpf@wrt}{\@prolpf@}}
\def\@prolpf@wrt[#1]#2{\@ifnextchar[{\@prolpf@wrt@at{#1}{#2}}{\@prolpf@wrt@{#1}{#2}}}

\def\@prolpf@wrt@at#1#2[#3]{\prolsymbol^{#1}_{#3}#2}
\def\@prolpf@wrt@#1#2{\prolsymbol^{#1}#2}
\def\@prolpf@#1{\@ifnextchar[{\@prolpf@at{#1}}{\@prolpf@@{#1}}}
\def\@prolpf@at#1[#2]{\prolsymbol_{#2}#1}
\def\@prolpf@@#1{\prolsymbol#1}
\def\@proldf{\@ifnextchar[{\@proldf@wrt}{\@proldf@}}
\def\@proldf@wrt[#1]#2{\@ifnextchar[{\@proldf@wrt@at{#1}{#2}}{\@proldf@wrt@{#1}{#2}}}

\def\@proldf@wrt@at#1#2[#3]{\prolsymbol^{*#1}_{#3}#2}
\def\@proldf@wrt@#1#2{\prolsymbol^{*#1}#2}
\def\@proldf@#1{\@ifnextchar[{\@proldf@at{#1}}{\@proldf@@{#1}}}
\def\@proldf@at#1[#2]{\prolsymbol^*_{#2}#1}
\def\@proldf@@#1{\prolsymbol^*#1}
\def\prolsymbol{\mathcal{L}}
\makeatother

\def\lcf{\lbrack\! \lbrack}
\def\rcf{\rbrack\! \rbrack}
\def\r{\ensuremath{\mathbb{R}}}
\def\rk{{\mathbb R}^{k}}
\def\tkq{T^1_kQ}
\def\tkqh{(T^1_k)^*Q}

\def\taukq{\tau^k_Q}

\def\ke{\stackrel{k}{\oplus} E}
\def\te{\mathcal{T}^E(\ke)}
\def\keh{\stackrel{k}{\oplus} {E^{\,*}}}
\def\teh{\mathcal{T}^E(\keh)}

\def\Le{\mathfrak{Leg}} 
\def\derpar#1#2{\ds\frac{\partial{#1}}{\partial{#2}}}
\def\derpars#1#2#3{\ds\frac{\partial^2{#1}}{\partial{#2}{\partial
{#3}}}}
\def\vf{\mathfrak{X}}
\def\d{{\rm d}}

\def\bea{\begin{eqnarray}}
\def\eea{\end{eqnarray}}
\def\beann{\begin{eqnarray*}}
\def\eeann{\end{eqnarray*}}
\def\beasn{\begin{sneqnarray}}
\def\eeasn{\end{sneqnarray}}
\def\ben{\begin{enumerate}}
\def\een{\end{enumerate}}
\def\bit{\begin{itemize}}
\def\eit{\end{itemize}}
\def\proof{ ({\sl Proof\/}) }
\def\derpar#1#2{\ds\frac{\partial{#1}}{\partial{#2}}}
\def\derpars#1#2#3{\ds\frac{\partial^2{#1}}{\partial{#2}{\partial
{#3}}}}

\def\qed{\ifvmode\Realemovelastskip\fi
{\unskip\nobreak\hfil\penalty50\hbox{}\nobreak\hfil \hbox{\vrule
height1.2ex width1.2ex}\parfillskip=0pt \finalhyphendemerits=0
\par\smallskip}}

\def\vf{\mathfrak{X}}

\def\d{{\rm d}}

\def\rk{\mathbb{R}^k}
\def\r{\mathbb{R}}
\def\tkq{T^1_kQ}

\def\Le{\mathfrak{Leg}}

\def\qed{\ifvmode\removelastskip\fi
{\unskip\nobreak\hfil\penalty50\hbox{}\nobreak\hfil \hbox{\vrule
height1.2ex width1.2ex}\parfillskip=0pt \finalhyphendemerits=0
\par\smallskip}}

\newcommand{\ds}{\displaystyle}

\begin{document}

\title{$k$-symplectic
formalism on Lie algebroids}

\author[M. de Le\'on]{M. de Le\'on}
\address{M. de Le\'on:
Instituto de Ciencias Matem\'aticas (CSIC-UAM-UC3M-UCM), Consejo
Superior de Investigaciones Cient\'{\i}ficas, Serrano 123, 28006
Madrid, Spain} \email{mdeleon@imaff.cfmac.csic.es}
\author[D.\ Mart\'{\i}n de Diego]{D. Mart\'{\i}n de Diego}
\address{D.\ Mart\'{\i}n de Diego:
Instituto de Ciencias Matem\'aticas (CSIC-UAM-UCM-UC3M)\\ C/ Serrano
123, 28006 Madrid, Spain} \email{d.martin@imaff.cfmac.csic.es}
\author[M. Salgado]{M. Salgado}
\address{Modesto Salgado:
Departamento de Xeometr\'{\i}a e Topolox\'{\i}a, Facultade de
Matem\'{a}ticas,
    Universidade de Santiago de Compostela,
    15782-Santiago de Compostela, Spain}
\email{modesto@zmat.usc.es}

\author[S. Vilari\~no]{S. Vilari\~no}
\address{Silvia Vilari\~no:
Departamento de Xeometr\'{\i}a e Topolox\'{\i}a, Facultade de
Matem\'{a}ticas,
    Universidade de Santiago de Compostela,
    15782-Santiago de Compostela, Spain}
    \email{silvia.vilarino@usc.es}



\begin{abstract}
 In this paper we introduce a geometric description of Lagrangian and
Hamiltonian classical field theories on Lie algebroids in the framework
of $k$-symplectic geometry. We discuss the relation between Lagrangian
and Hamiltonian descriptions through a convenient notion of Legendre
transformation. The theory is a natural generalization of the standard
one; in addition, other interesting examples are studied, in particular,
systems with  symmetry and Poisson sigma models.

\end{abstract}


\maketitle

\begin{center}\today\end{center}

\tableofcontents

\section{Introduction}\label{intro}

The notion of Lie algebroid is a generalization of both the concept of a Lie algebra and the concept of an integrable distribution.
The idea of using Lie algebroids in Mechanics is due to A. Weinstein \cite{Weins-1996}. He introduced a new geometric framework
for the  description of Lagrangian Mechanics. His formulation allows us to describe geometrically, in a unified way, different types  of dynamical
systems; such as those whose lagrangian systems whose phase spaces are Lie groups, Lie algebras,
cartesian products of manifolds or quotient manifolds
(as it happens, for instance, in the reduction theory,
where the reduced phase spaces are not, in general, tangent or cotangent bundles).
This approach was followed and completed by other authors in order to study
different kinds of problems concerning mechanical systems (a good
survey on this subject is  \cite{LMM-2005}).

In this paper we will study an extension of mechanics on Lie algebroids to classical field theories. Classical field theories on Lie algebroids have already been studied in the literature. For instance, the multisymplectic formalism on Lie algebroids was presented in \cite{Mart,{Mart-2005}}. In \cite{VC-06} a geometric framework for discrete field theories on
Lie groupoids has been discussed.

 The
multisymplectic formalism was developed by Tulczyjews school in
Warsaw (see, for instance, \cite{KT}), and independently by Garc\'{\i}a and
P\'{e}rez-Rend\'{o}n \cite{PLG1,PLG2} and Goldschmidt and Sternberg \cite{GS}. This
approach was revised, among others, by Martin \cite{mar,mar2} and Gotay et al \cite{GIMMSY-mm} and, more recently, by Cantrijn et al \cite{CIL1}.

An alternative way to derive  certain types of the field equations
is to use the G\"{u}nther ($k$-symplectic) formalism. The $k$-symplectic formalism
is the generalization to field theories of the standard symplectic
formalism in Mechanics, which is the geometric framework for
describing autonomous dynamical systems. In this sense, the
$k$-symplectic formalism is used to give a geometric description of
certain kinds of field theories: in a local description, those
theories whose Lagrangian and Hamiltonian do not depend on the base coordinates,
denoted by $(t^1, \ldots , t^k)$ (in many of the cases defining the space-time
coordinates); that is, the $k$-symplectic formalism is only valid for
Lagrangians $L(q^i, v^i_A )$ and Hamiltonians $H(q^i, p^A_i )$ that
depend on the field coordinates $q^i$ and on the partial derivatives
of the field $v^i_A$, or the corresponding moment $p^A_i$. To treat with that general situation we need to extend the formalism using $k$-cosymplectic geometry, see \cite{LMORS-1998, LMS-2001}.

G\"{u}nther's paper  \cite{Gu-1987} gave a geometric Hamiltonian formalism
for field theories. The crucial device is the introduction of a
vector-valued generalization of a symplectic form, called a
polysymplectic form. One of the advantages of this formalism is that
one only needs the tangent and cotangent bundle of a manifold to
develop it. In \cite{MRS-2004} G\"{u}nther's formalism has been revised
and clarified. It has been shown that the polysymplectic structures
used by G\"{u}nther to develop his formalism could  be replaced by the
$k$-symplectic structures defined independently by Awane \cite{Aw-1992,{Aw-2000}}, L. K. Norris \cite{McN,No2,No3,No4,No5} and de Leon {\it et al.} \cite{LMS-1988,{LMS-1993}}. So this
formalism is also called $k$-symplectic formalism (see also
\cite{mt1,mt2}).

The purpose of this paper is to give a $k$-symplectic setting to first-order classical field theories on Lie algebroids. In the $k$-symplectic setting we will present a geometric description of Lagrangian and Hamiltonian classical field theories on Lie algebroids and we will find the relation between the solutions of both formalism when the Lagrangian is hyperregular.

The organization of the paper is as follows. In section 2 we recall some basic elements from the $k$-symplectic approach to first order classical field theories. In section 3 we recall some basic facts about Lie algebroids an the differential geometry associated to them. In this section we also describe a particular example of Lie algebroid, called the {\it prolongation of a Lie algebroid over a fibration}. This Lie algebroid will be necessary for the further developments. In section 4 the $k$-symplectic formalism is extended to the setting of Lie algebroids. The subsection 4.1 describe the Lagrangian approach and the subsection 4.2 describe the Hamiltonian approach. These formalism are developed in an analogous way to the standard $k$-symplectic Lagrangian and Hamiltonian formalism. We finish this section defining the Legendre transformation on the context of Lie algebroids and we establish the equivalence between both formalism, Lagrangian and Hamiltonian, when the Lagrangian function is hyperregular. In section 5 we show some examples where the theory can be applied to the Poisson-Sigma model or first order field theories with symmetries.

 All manifolds and maps are $C^\infty$. Sum over crossed repeated
indices is understood. Along this paper one $k$-tuple of elements will be denoted by a bold symbol.

\section{Geometric preliminaries}\label{Geprel}
 In this section  we recall some basic
elements from the $k$-symplectic approach to   classical field
theories. The contents of this section can be found in
\cite{Gu-1987,MRS-2004,RSV-2007}.
\subsection{The tangent bundle of $k^1$-velocities of a manifold}
Let $\tau_Q : TQ \to Q$ be the tangent bundle of $Q$, where $Q$ is as $n$-dimensional differentiable manifold. Let us denote
by $T^1_kQ$ the Whitney sum $TQ \oplus \stackrel{k}{\dots} \oplus
TQ$ of $k$ copies of $TQ$, with projection $\tau^k_Q : T^1_kQ \to
Q$, $\tau^k_Q ({v_1}_ {q},\ldots ,
{v_k}_ {q})= {q}$, where ${v_A}_{q}\in T_{q}Q,\, A=1,\ldots, k$.

$T^1_kQ$ can be identified with the manifold $J^1_{\mathbf{0}}(\r^k,Q)$ of
{\it $k^1$-velocities    of $Q$}, that is,  $1$-jets of maps
$\sigma:\rk\to Q$  with source at $\mathbf{0}\in \r^k$, say
\[
\begin{array}{ccc}
J^1_{\mathbf{0}}(\r^k,Q) & \equiv & TQ \oplus \stackrel{k}{\dots} \oplus TQ \\
j^1_{\mathbf{0}, {q}}\sigma & \equiv & ({v_1}_ {q},\ldots ,
{v_k}_ {q})
\end{array}
\]
where $ {q}=\sigma (\mathbf{0})$,  and ${v_A}_ {q}=
\sigma_*(\mathbf{{0}})(\ds\frac{\partial}{\partial t^A}\Big\vert_{\mathbf{{0}}})$. Here $(t^1,\ldots, t^k)$ denote the standard coordinates on $\R^k$.
$T^1_kQ$ is called  {\it the tangent bundle of $k^1$-velocities of
$Q$} (see \cite{mor}).

If $(q^i)$ are local coordinates on $U \subseteq Q$ then the induced
local coordinates $(q^i , v^i)$, $1\leq i \leq n$, on
$TU=\tau_Q^{-1}(U)$ are expressed by
$$ q^i(v_ {q})=q^i( {q}),\qquad
  v^i(v_ {q})=v_ {q}(q^i)  $$
and  the induced local coordinates $(q^i , v_A^i)$, $1\leq i \leq
n,\, 1\leq A \leq k$, on $T^1_kU=(\tau^k_Q)^{-1}(U)$ are given by
$$ q^i({v_1}_ {q},\ldots , {v_k}_ {q})=q^i( {q}),\qquad
  v_A^i({v_1}_ {q},\ldots , {v_k}_ {q})={v_A}_ {q}(q^i) \, .$$

 Let $f:M \to N$ be a differentiable
map, then the induced map $T^1_kf:T^1_kM \to  T^1_kN$  defined by
$T^1_kf(j^1_{\mathbf{0}}\sigma)=j^1_{\mathbf{0}}(f \circ \sigma)$ is called the {\it
canonical prolongation} of $f$. Observe that from its definition:
$$  T^1_kf({v_1}_ {q},\ldots , {v_k}_ {q})=(f_*( {q})({v_1}_ {q}),\ldots
,f_*( {q})({v_k}_ {q})) \quad ,$$ where
${v_1}_ {q},\ldots , {v_k}_ {q}\in T_ {q}Q$, $ {q}\in
Q$.

\subsection{$k$-vector fields and integral sections}  Let
$M$ be an arbitrary manifold.
\begin{definition}  \label{kvector} A section
$\mathbf{X} : M \longrightarrow T^1_kM$ of the projection $\tau^k_M$
will be called a {\rm $k$-vector field} on $M$. \end{definition}

 Since
$T^{1}_{k}M$ is  the Whitney sum $TM\oplus \stackrel{k}{\dots}
\oplus TM$ of $k$ copies of $TM$,
  we deduce that to give a $k$-vector field $\mathbf{X}$ is equivalent to give
 a family of $k$ vector fields $X_{1}, \dots, X_{k}$ on $M$ obtained by
projecting $\mathbf{X}$ on each factor. For this reason we will
denote a $k$-vector field by $\mathbf{X}=(X_1, \ldots, X_k)$.

\begin{definition}
\label{integsect} An {\rm integral section}  of the $k$-vector field
$\mathbf{X}=(X_{1}, \dots,X_{k})$, passing through a point
$ {x}\in M$, is a map $\psi\colon U_{\mathbf{0}}\subset \r^k \rightarrow
M$, defined on some neighborhood  $U_{\mathbf{0}}$ of ${\mathbf{0}}\in \r^k$,  such that
$$
\psi({\mathbf{0}})= {x}, \, \, \psi_{*}(\mathbf{t})\left(\frac{\partial}{\partial
t^A}\Big\vert_\mathbf{t}\right)=X_{A}(\psi (\mathbf{t}))
 \; , \quad \mbox{\rm for every $\mathbf{t}\in U_{ {\mathbf{0}}}$, $1\leq A \leq k$}
$$
or,  equivalently,  $\psi$ satisfies that
${  \mathbf{X}}\circ\psi=\psi^{(1)}$, where  $\psi^{(1)}$ is the first
prolongation of $\psi$  to $T^1_kM$ defined by
$$
\begin{array}{rccl}\label{1prolong}
\psi^{(1)}: & U_{0}\subset \r^k & \longrightarrow & T^1_kM
\\\noalign{\medskip}
 &\mathbf{t} & \longrightarrow & \psi^{(1)}(\mathbf{t})=j^1_{\mathbf{0}}\psi_\mathbf{t}\equiv
 \left(\psi_*(\mathbf{t})\left(\derpar{}{t^1}\Big\vert_\mathbf{t}\right),\ldots,
\psi_*(\mathbf{t})\left(\derpar{}{t^k}\Big\vert_\mathbf{t}\right)\right) \, ,
 \end{array}
$$ where $\psi_\mathbf{t}({  \mathbf{s}})=\psi(\mathbf{t}+{  \mathbf{s}})$.

 A $k$-vector field ${  \mathbf{X}}=(X_1,\ldots , X_k)$ on $M$ is said to
 be {\rm integrable} if there is an integral section passing through
every point of $M$.
\end{definition}
\begin{remark}{\rm In the $k$-symplectic formalism,  the solutions of the field equations are described as the integral
 sections of some $k$-vector fields. Observe that, in the case $k=1$, this definition coincides with
the classical definition of integral curve of a vector field. }\end{remark}

In  a local coordinate system, if $\psi(\mathbf{t})=(\psi^i(\mathbf{t}))$ then one
has
\begin{equation}
\label{localfi11} \psi^{(1)}( \mathbf{t})=\left( \psi^i ( \mathbf{t}), \frac{\partial\psi^i}{\partial t^A}\Big\vert_{ \mathbf{t}}\right), \qquad  1\leq A\leq k\, ,\, 1\leq i\leq n \, ,
\end{equation}
and $\psi$ is an integral section of $(X_1,\ldots, X_k)$ if and only
if the following  equations holds:
\begin{equation}\label{intsect}\derpar{\psi^i}{t^A}= X_A^i\circ\psi\, \quad 1\leq A\leq k,\;
1\leq i\leq n\;,\end{equation} being $X_A=X_A^i\ds\frac{\partial}{\partial q^i}$.

\subsection{The cotangent bundle of $k^1$-covelocities of a
manifold}

 Let $Q$ be a differentiable manifold of dimension  $n$
and $\pi_Q: T^*Q \to Q$ its cotangent bundle. Denote by
$(T^1_k)^*Q= T^*Q\oplus \stackrel{k}{\dots} \oplus T^*Q$ the Whitney
sum of  $T^*Q$ with itself $k$ times, with projection map $\pi^k_Q\colon
(T^1_k)^* Q \to Q$, $\pi^k_Q (\alpha_{1_ {q}},\ldots
,\alpha_{k_ {q}})= {q}$.

Observe that the manifold  $(T^1_k)^*Q$ can be canonically identified with the
vector bundle $J^1(Q,\r^k)_{\mathbf{0}}$ of $k^1$-covelocities of the manifold
$Q$, the manifold of $1$-jets of maps $\sigma\colon Q\to\r^k$ with
target at ${\mathbf{0}}\in \r^k$ and projection map $\pi^k_Q\colon
J^1(Q,\r^k)_{\mathbf{0}}\to Q$, $\pi^k_Q
(j^1_{ {q},{\mathbf{0}}}\sigma)= {q}$; that is,
\[
\begin{array}{ccc}
J^1(Q,\r^k)_{\mathbf{0}} & \equiv & T^*Q \oplus \stackrel{k}{\dots} \oplus T^*Q \\
j^1_{ {q},{\mathbf{0}}}\sigma & \equiv & (d\sigma_1( {q}), \dots
,d\sigma_k( {q}))
\end{array}
\]
where $\sigma_A= pr_A \circ \sigma:Q \longrightarrow \r$ is the
$A$-th component of $\sigma$, and  $pr_A\colon \r^k \to \r$ are the
canonical projections, $1\leq A \leq k$. For this reason,
$(T^1_k)^*Q$ is also called
 {\sl the bundle of $k^1$-covelocities of the manifold $Q$.}

If $(q^i)$ are local coordinates on $U \subseteq Q$,  then the
induced local coordinates $(q^i , p_i)$, $1\leq i \leq n$, on
$T^*U=(\pi_Q)^{-1}(U)$,   are given by
$$
q^i(\alpha_ {q})=q^i( {q}), \quad
p_i(\alpha_ {q})=
\alpha_ {q}\left(\frac{\partial}{\partial
q^i}\Big\vert_ {q}\right)
$$
and the induced  local coordinates $(q^i , p^A_i),\, 1\leq i \leq
n,\, 1\leq A \leq k$, on $(T^1_k)^*U=(\pi^k_Q)^{-1}(U)$ are
$$
  q^i(\alpha_{1_ {q}},\ldots , \alpha_{k_ {q}})=q^i( {q}),\qquad
p^A_i(\alpha_{1_ {q}},\ldots , \alpha_{k_ {q}})=
\alpha_{A_ {q}}\left(\frac{\partial}{\partial
q^i}\Big\vert_ {q}\right)\, .
$$

We can endow $(T^1_k)^*Q$ with a $k$-symplectic structure  given by
the family $(\omega^1,\ldots, $ $\omega^k; V=\ker T\pi^k_Q)$ where
each $\omega^A$ is the $2$-form given by
$$
\omega^A = (\pi_Q^{k,A})^*\omega_Q, \quad 1\leq A \leq k\, ,
$$
being  $\pi_Q^{k,A}\colon (T^1_k)^*Q \rightarrow T^*Q $ the
canonical projection onto the $A^{th}$-copy  $T^*Q$ of $(T^1_k)^*Q$
and $\omega_Q$ is the canonical sympletic form on $T^*Q$. Therefore, in local coordinates, we have $\omega^A= dq^i\wedge dp^A_i$. (See
\cite{Aw-1992,Aw-2000,MRS-2004,RSV-2007})

\section{Lie algebroids}\label{algebroids}

In this section we present some basic facts on Lie algebroids,
including results from the associated differential calculus and Lie algebroids morphisms, that will be
necessary for the further developments. We refer the reader to
\cite{CW-1999,HM-1990,Mack-1987,Mack-1995} for details about Lie
groupoids, Lie algebroids and their role in differential geometry.

\subsection{Lie algebroid: definition}

Let $E$ be a vector bundle of rank $m$ over a manifold $Q$ of
dimension $n$ and $\tau: E\to Q$ be the vector bundle projection.
Denoted by ${\rm Sec}(E)$ the $C^\infty(Q)$-module of sections of
$\tau: E\to Q$. A {\it Lie algebroid structure}
$(\lcf\cdot,\cdot\rcf_E,\rho_E)$ on $E$ is a Lie bracket
$\lcf\cdot,\cdot\rcf_E: {\rm Sec}(E)\times {\rm Sec}(E)\to {\rm Sec}(E)$ on
the space ${\rm Sec}(E)$, together with  a bundle  map $\rho_E: E\to
TQ$, called {\it the anchor map}, such that if we also denote by
$\rho_E:{\rm Sec}(E)\to \vf{(Q)}$ the homomorphism of the
$C^\infty(Q)$-modules induced by the anchor map then it is satisfied
the following {\it compatibility condition}
\[
\lcf\sigma_1,f\sigma_2\rcf_E=
f\lcf\sigma_1,\sigma_2\rcf_E+(\rho_E(\sigma_1)f)\sigma_2\,.\]

Here $f$ is a smooth function on $Q$,$\,\sigma_1,\sigma_2$ are
sections of $E$ and we have denoted by $\rho_E(\sigma_1)$ the vector
field on $Q$ given by
$\rho_E(\sigma_1)( {q})=\rho_E(\sigma_1( {q}))$. The triple
$(E,\lcf\cdot,\cdot\rcf_E,\rho_E)$ is called a {\it Lie algebroid over $Q$}.
{}From the compatibility condition and the Jacobi identity, it follows
that the anchor map $\rho_E:{\rm Sec}(E)\to\vf{(Q)}$ is a homomorphism
between the Lie algebras $({\rm Sec}(E),\lcf\cdot,\cdot\rcf_E)$ and
$(\vf{(Q)},[\cdot,\cdot])$.

Some examples of Lie algebroids over $Q$ are:
\begin{enumerate}
\item {\bf Real Lie algebras of finite dimension}. Let
$\mathfrak{g}$ be a real Lie algebra of finite dimension. Then, it
is clear that $\mathfrak{g}$ is a Lie algebroid over a single point.

\item {\bf The tangent bundle.} Let $TQ$ be the tangent bundle of a
manifold $Q$. Then, the triple $(TQ,[\cdot,\cdot],id_{TQ})$ is a Lie
algebroid over $Q$, where $id_{TQ}:TQ\to TQ$ is the identity map.

\item Another interesting example of a Lie algebroid may be constructed as follows.
Let $\pi: P\to Q$ be a principal bundle with structural group $G$.
Denote by $\Phi:G\times P\to P$ the free action of $G$ on $P$ and
by $T\Phi:G\times TP\to TP$ the tangent action of $G$ on $TP$.
Then, one may consider the quotient vector bundle
$\tau_{P|G}: TP/G\to Q=P/G$ and the sections of this vector bundle
may be identified with the vector fields on $P$ which are
invariant under the action $\Phi$. Using that every $G$-invariant
vector field on $P$ is $\pi$-projectable and the fact that the
standard Lie bracket on vector fields is closed with respect to
$G$-invariant vector fields, we can induce a Lie algebroid
structure on $TP/G$. The resultant Lie algebroid is called
{\bf the Atiyah (gauge) algebroid associated with the principal
bundle} $\pi:P\to Q$ (see \cite{LMM-2005,Mack-1987}).

\end{enumerate}

 Along this paper, the Lie algebroid will play the role of a substitute of the tangent bundle of
$Q$. In this way, one regards an element $e$ of $E$ as a generalized
velocity, and the actual velocity ${  v}$ is obtained when we apply the
anchor map to ${  e}$, i.e. ${  v}=\rho_E({  e})$.

Let $(q^i)_{i=1}^n$ be local coordinates on $Q$ and
$\{e_\alpha\}_{1\leq \alpha\leq m}$ be a local basis of sections of $\tau$.
Given ${  e}\in E$ such that $\tau({  e})= {q}$, we can write
$e=y^\alpha(e)e_\alpha( {q})\in E_ {q}$, thus the
coordinates of ${  e}$ are $(q^i({  e}),y^\alpha({  e}))$. Therefore, each
section $\sigma$ is locally given by $\sigma\big\vert_{U}=y^\alpha
e_\alpha$.

In local form, the Lie algebroid structure is determined by a set of
local functions  $\rho^i_\alpha,\; \mathcal{C}^\gamma_{\alpha\beta}$
on $Q$. They are determined by the relations
\begin{equation}\label{structure} \rho_E(e_\alpha)=\rho^i_\alpha
\ds\frac{\partial}{\partial q^i} ,\quad
\lcf e_\alpha,e_\beta\rcf_E=\mathcal{C}^\gamma_{\alpha\, \beta}e_\gamma\, .
\end{equation} The functions $\rho^i_\alpha$ and
$\mathcal{C}^\gamma_{\alpha\beta}$ are called the {\it structure
functions} of the Lie algebroid in the above coordinate system. They
satisfy the following relations (as a consequence of the compatibility
condition and Jacobi's identity):
\begin{equation}\label{ecest}\ds\sum_{cyclic(\alpha,\beta,\gamma)}\left(\rho^i_\alpha\ds\frac{\partial
\mathcal{C}^\nu_{\beta\gamma}}{\partial
q^i}+\mathcal{C}^\nu_{\alpha\mu}\mathcal{C}^\mu_{\beta\gamma}\right)=0\;
\quad , \quad  \rho^j_\alpha\ds\frac{\partial \rho^i_\beta}{\partial
q^j}- \rho^j_\beta\ds\frac{\partial \rho^i_\alpha}{\partial
q^j}=\rho^i_\gamma \mathcal{C}^\gamma_{\alpha\beta}\;,
\end{equation}which are usually called {\it the structure equations} of the Lie algebroid $E$.

\subsection{Exterior differential}
The structure of Lie algebroid on
$E$ allows us to define {\it the exterior differential of $E$},
$d^E:{\rm Sec}(\bigwedge^l E^*)\to {\rm Sec}(\bigwedge^{l+1} E^*)$,
as follows
{\small\begin{equation}\label{difE}\begin{array}{lcl}
d^E\mu(\sigma_1,\ldots, \sigma_{l+1})&=&
\ds\sum_{i=1}^{l+1}(-1)^{i+1}\rho_E(\sigma_i)\mu(\sigma_1,\ldots,
\widehat{\sigma_i},\ldots, \sigma_{l+1})\\\noalign{\medskip} &+&
\ds\sum_{i<j}(-1)^{i+j}\mu([\sigma_i,\sigma_j]_E,\sigma_1,\ldots,
\widehat{\sigma_i},\ldots, \widehat{\sigma_j},\ldots
\sigma_{l+1})\;,
\end{array}\end{equation}}\noindent for $\mu\in {\rm Sec}(\bigwedge^lE^*)$ and
$\sigma_1,\ldots,\sigma_{l+1}\in {\rm Sec}(E)$. It follows that $d^E$
is a cohomology operator, that is, $(d^E)^2=0$.

In particular, if $f:Q\to\r$ is a real smooth function then
$d^E f(\sigma)=\rho_E(\sigma)f$, for $\sigma\in {\rm Sec}(E)$. Locally, the
exterior differential is determined by
\[d^Eq^i=\rho^i_\alpha e^\alpha\quad \makebox{and}\quad
d^Ee^\gamma=-\ds\frac{1}{2}\mathcal{C}^\gamma_{\alpha\beta}e^\alpha\wedge
e^\beta\,,\]where $\{e^\alpha\}$ is the dual basis of
$\{e_\alpha\}$.

The usual Cartan calculus extends to the case of Lie algebroids: for
every section $\sigma$ of $E$ we  have a derivation $\imath_\sigma$
(contraction) of degree $-1$ and a derivation
$\mathcal{L}_\sigma=\imath_\sigma\circ d + d\circ \imath_\sigma$
(Lie derivative) of degree $0$ (for more details, see
\cite{Mack-1987,Mack-1995}).

\subsection{Morphisms}
 Let $(E,\lcf\cdot,\cdot\rcf_E,\rho_E)$ and
$(E',\lcf\cdot,\cdot\rcf_{E'},\rho_{E'})$ be two Lie algebroids over $Q$ and
$Q'$ respectively, and suppose that
$\Phi=(\overline{\Phi},\underline{\Phi})$ is a vector bundle map,
that is $\overline{\Phi}:E\to E'$ is a fiberwise linear map over
$\underline{\Phi}:Q\to Q'$. The pair
$(\overline{\Phi},\underline{\Phi})$ is said to be a {\it Lie
algebroid morphism} if \begin{equation}\label{lie morph}d^E
(\Phi^*\sigma')=\Phi^*(d^{E'}\sigma ')\,,\quad\makebox{ for all }
\sigma '\in Sec(\textstyle\bigwedge^l (E')^*)\makebox{ and for all
$l$.}\end{equation}

Here $\Phi^*\sigma '$ is the section of the vector
bundle $\bigwedge^l E^*\to Q$ defined (for $l>0$) by
\begin{equation}\label{pullsec}(\Phi^*\sigma ')_ {q}({  e}_1,\ldots,{  e}_l) = \sigma_{\underline{\Phi}
( {q})}' (\overline{\Phi}({  e}_1),\ldots,
\overline{\Phi}({  e}_l))\,,\end{equation} for $ {q}\in Q$ and ${  e}_1,\ldots,
{  e}_l\in E_ {q}$.  In  particular   when $Q=Q'$ and
$\underline{\Phi}=id_Q$ then (\ref{lie morph}) holds if and only if
\[\lcf\overline{\Phi}\circ\sigma_1,\overline{\Phi}\circ\sigma_2\rcf_{E'} =
\overline{\Phi}\lcf\sigma_1,\sigma_2\rcf_E,\quad
\rho_{E'}(\overline{\Phi}\circ\sigma)=\rho_E(\sigma),\quad \makebox{for }
\sigma,\sigma_1,\sigma_2\in Sec(E)\,.\]

Let $(q^i)$ be a local coordinate system on $Q$ and $(\bar{q}^i)$ a
local coordinate system on $Q'$. Let $\{e_\alpha\}$ and
$\{\bar{e}_{\bar{\alpha}}\}$ be a local basis of section of $E$ and $E'$,
respectively, and $\{e^\alpha\}$ and $\{\bar{e}^{\bar{\alpha}}\}$ their dual
basis, respectively. The vector bundle map $\Phi$ is determined by the relations
$\Phi^*{\bar{q}^{\bar{i}}}=\phi^{\bar{i}}( {q})$ and $\Phi^* \bar{e}^{\bar{\alpha}}=
\phi^{\bar{\alpha}}_\beta e^\beta$ for certain local functions $\phi^{\bar{i}}$ and
$\phi^{\bar{\alpha}}_\beta$ on $Q$. In this coordinate system
$\Phi=(\overline{\Phi},\underline{\Phi})$ is a   Lie
algebroids morphism if and only if
\begin{equation}\label{morp cond}
  (\rho_E)^j_\alpha \ds\frac{\partial \phi^{\bar{i}}}{\partial q^j} =
  (\rho_{E'})^{\bar{i}}_{\bar{\beta}}\phi^{\bar{\beta}}_\alpha\quad , \quad
\phi^{\bar{\beta}}_\gamma\mathcal{C}^\gamma_{\alpha\delta}
 =\left((\rho_E)^i_\alpha\ds\frac{\partial \phi^{\bar{\beta}}_\delta}{\partial
q^i} - (\rho_E)^i_\delta \ds\frac{\partial \phi^{\bar\beta}_\alpha}{\partial
q^i}\right) +
{\bar{\mathcal{C}}}^{\bar{\beta}}_{\bar{\theta}\bar{\sigma}}\phi^{\bar{\theta}}_\alpha\phi^{\bar{\sigma}}_\delta
\,.\end{equation} In these expressions $(\rho_E)^i_\alpha,
\mathcal{C}^\alpha_{\beta\gamma}$ are the structure functions on $E$,
and $(\rho_{E'})^{\bar{i}}_{\bar{\alpha}}, {\bar{\mathcal{C}}}^{\bar\alpha}_{\bar\beta\bar\gamma}$ are the
structure functions on $E'$.

\subsection{The prolongation of a Lie algebroid over a fibration}\label{prolong}
(See \cite{CLMMM-2006,HM-1990,LMM-2005,Mart-2001}). In this subsection we describe a particular example of Lie algebroid which will be necessary for the further developments.

Let
$(E,\lcf\cdot , \cdot\rcf_E,\rho_E)$ be a Lie algebroid over a manifold $Q$
and $\pi:P\to Q$ be a fibration. We consider the subset of $E\times
TP$
\[\mathcal{T}^E_{  p}P=\{( {e}, {v}_{  {p}})\in E_{  q}\times T_{  p}P\, |\,
\rho_E( {e})=T_{{  p}}\pi( {v}_{  {p}})\}\;,\] where $T\pi: TP\to TQ$ is the
tangent map to $\pi\,,\, {  p}\in P $ and
$\pi({  p})={  q}$.
\[\,\mathcal{T}^EP=\ds\bigcup_{{  p}\in
P}\mathcal{T}^E_{  p}P\] is a vector bundle over $P$ with projection $\widetilde{\tau}_P:\mathcal{T}^EP\to P$   given by
\[
\widetilde{\tau}_P( {e}, {v}_{ {p}})=\tau_P( {v}_{ {p}})= {p}\,,
\]
being $\tau_P\colon TP\to P$ the canonical projection.

Next, we will see that it is possible to induce a Lie algebroid structure on $\widetilde{\tau}_P\colon \mathcal{T}^EP\to P$. The anchor map $\rho^{\pi}$ is given as follows: $\rho^\pi:\mathcal{T}^EP\to TP,\;\rho^\pi( {e}, {v}_{  p})= {v}_{  p}$.

In order to introduce a Lie bracket on ${\rm Sec}(\mathcal{T}^EP)$, the set of sections of $\widetilde{\tau}_P$, we first consider a local basis of ${\rm Sec}(\mathcal{T}^EP)$.

Given local coordinates $(q^i,u^\ell)$ on $P$ and a local basis $\{e_\alpha\}$ of sections of $E$, we can define a local basis $\{\mathcal{X}_\alpha,\mathcal{V}_\ell\}$ of sections of $\widetilde{\tau}_P\colon \mathcal{T}^EP\to P$ by
\begin{equation}\label{base k-prol}
  \mathcal{X}_\alpha( {p}) =
(e_\alpha(\pi( {p}));\rho^i_\alpha(\pi( {p}))\ds\frac{\partial
}{\partial q^i}\Big\vert_ {p}) \quad \makebox{and}\quad
 \mathcal{V}_\ell( {p})
=({ {0}}_{\pi( {p})};\ds\frac{\partial}{\partial
u^\ell}\Big\vert_ {p})\,.
\end{equation}

If $ {z}=( {e}, {v}_{ {p}})$ is an element of $\mathcal{T}^EP$, with $ {e}=z^\alpha e_\alpha$, then $ {v}_{ {p}}$ is of the form $ {v}_{ {p}}=\rho^i_{\alpha}z^\alpha\frac{\partial}{\partial q^i}\Big\vert_ {p} + v^\ell\frac{\partial}{\partial u^\ell}\Big\vert_ {p}$, and we can write
\[
 {z}=z^\alpha\mathcal{X}_{\alpha}( {p}) + v^\ell\mathcal{V}_\ell( {p})\,.
\]

The anchor map $\rho^{\pi}$ applied to a section $Z$ of $\mathcal{T}^EP$ with local expression $Z= Z^\alpha\mathcal{X}_\alpha + V^\ell\mathcal{V}_\ell$ is the vector field on $P$ whose coordinate expression is
\begin{equation}\label{rhoz}\rho^{\pi}(Z)= \rho^i_\alpha Z^\alpha\derpar{}{q^i} +
V^\ell\derpar{}{u^\ell}\in \vf(P)\,.\end{equation}

Now, we will introduce a Lie bracket structure on the space of sections of $\mathcal{T}^EP$. For that, we say that a section $Z$ of $\mathcal{T}^EP$ is {\it projectable} if there exists a section $\sigma$ of $\tau\colon E\to Q$ and a vector field $X\in\mathfrak{X}(P)$, which is $\pi$-projectable to the vector field $\rho(\sigma)$ and such that $Z( {p})=(\sigma(\pi( {p})), X( {p})),$ for all $ {p}\in P$. For such a projectable section $Z$, we will use the following notation $Z\equiv(\sigma,X)$. It is easy to prove that one may choose a local basis of projectable sections of the space ${\rm Sec}(\mathcal{T}^EP)$.

The Lie bracket of two projectable sections $Z=(\sigma, X)$ and $Z'=(\sigma',X')$ is then given by
\begin{equation}\label{Lie brack}\lcf Z,Z'\rcf^\pi( {p})
=(\lcf\sigma,\sigma'\rcf_E( {q}),[X,X']( {p})), \quad  {p}\in P, \; {q}=\pi( {p})\,.\end{equation}
Since any section of $\mathcal{T}^EP$ can be locally  written as a linear combination of projectable sections, the definition of the Lie bracket for arbitrary sections of $\mathcal{T}^EP$ follows. In particular, the Lie bracket of the elements of the local basis  $\{\mathcal{X}_\alpha,\mathcal{V}_\ell\}$ of ${\rm Sec}(\mathcal{T}^EP)$ is characterized by the following expressions
\begin{equation}\label{lie brack k-prol}\begin{array}{lll}
\lcf\mathcal{X}_\alpha,\mathcal{X}_\beta\rcf^{\pi}=
\mathcal{C}^\gamma_{\alpha\beta}\mathcal{X}_\gamma &
\lcf\mathcal{X}_\alpha,\mathcal{V}_\ell\rcf^{\pi}=0 &
\lcf\mathcal{V}_\ell,\mathcal{V}_\varphi\rcf^{\pi}=0\,,
\end{array}\end{equation}
and, therefore, the exterior differential is determined by
\begin{equation}\label{dtp}\begin{array}{lclclcl} d^{\mathcal{T}^EP}q^i &=&
\rho^i_\alpha\mathcal{X}^\alpha\,,&\qquad & d^{\mathcal{T}^EP}u^\ell
&=&
\mathcal{V}^\ell\\
d^{\mathcal{T}^EP}\mathcal{X}^\gamma &=&
-\ds\frac{1}{2}\mathcal{C}^\gamma_{\alpha\beta}
\mathcal{X}^\alpha\wedge\mathcal{X}^\beta\;,&\qquad &
d^{\mathcal{T}^EP}\mathcal{V}^\ell&=&0\end{array}\end{equation}
where $\{\mathcal{X}^\alpha,\mathcal{V}^\ell\}$ is the dual basis to
$\{\mathcal{X}_\alpha,\mathcal{V}_\ell\}$.

 The Lie algebroid $\mathcal{T}^EP$ is called the {\it prolongation of the Lie algebroid $E$ over the fibration $\pi\colon P\to Q$.} This Lie algebroid is very important in the $k$-symplectic formalism on Lie algebroids as we will see in the following section.

\section{Classical Field Theories on Lie algebroids: $k$-symplectic
approach}\label{CFTLA}

In this section, the $k$-symplectic formalism for first order classical field theories (see \cite{Gu-1987,MRS-2004,RSV-2007}) is
extended to the setting of Lie algebroids. Thinking on a Lie
algebroid $E$ as a generalization of the tangent bundle of $Q$, we define
the analog of the concept of field solution to the field equations
and we study the analog of the geometric structures of the standard
$k$-symplectic formalism.

In this section we will develop the Lagrangian and Hamiltonian
$k$-symplectic formalism on Lie algebroids (see subsections
\ref{Lagform} and \ref{Hamform}). Moreover, we also describe the
standard Lagrangian and Hamiltonian $k$-symplectic formalism as a
particular case of the formalism  developed here.

 Along this section
we consider a Lie algebroid $(E,\lcf\cdot,\cdot\rcf_E,\rho_E)$ on the
manifold $Q$. We note this Lie algebroid by $E$.
\subsection{Lagrangian formalism.}\label{Lagform}

\subsubsection{The manifold $\stackrel{k}{\oplus} E$}

The standard $k$-symplectic Lagrangian formalism is developed on the
bundle of $k^1$-velocities of $Q$, $T^1_kQ$, that is the Whitney sum
of $k$ copies of $TQ$. Since we are thinking  on a Lie algebroid $E$ as a
substitute of the tangent bundle, its natural to think that in this
situation, the analog of the bundle of $k^1$-velocities $T^1_kQ$ is
the Whitney sum of $k$ copies of the algebroid $E$.

We denote by $$\stackrel{k}{\oplus} E=E\oplus \stackrel{k}{\ldots}
\oplus E,$$ the Whitney sum of $k$ copies of the vector bundle $E$,
with projection map $$\widetilde{\tau} :\stackrel{k}{\oplus} E\to
Q,$$
given by $ \widetilde{\tau}  ( {e}_{1_ {q}},\ldots, {e}_{k_ {q}})=  {q}$. 

If $(q^i,y^\alpha)$ are local coordinates on
${\tau}^{-1}(U)\subseteq E$, then the induced local coordinates
$(q^i,y^\alpha_A)$ on $\widetilde{\tau}^{-1}(U)\subseteq
\stackrel{k}{\oplus} E$ are given by
$$
q^i( {e}_{1_ {q}},\ldots, {e}_{k_ {q}})=q^i( {q})\,,\quad
y^\alpha_A( {e}_{1_ {q}},\ldots, {e}_{k_ {q}})=y^\alpha( {e}_{A_ {q}})\;.
$$

\begin{remark} Consider the standard case where $E=TQ,\; \rho_{TQ}= id_{TQ}$. If we fix local coordinates
$(q^i)$ on $Q$, then we have the natural basis
 of ${\rm Sec}(TQ)=\vf(Q)$ given by  $\left\{\ds\frac{\partial}{\partial q^i}\right\}_{i=1}^n$. For this basis of section, obviously we have that
$\mathcal{C}^\gamma_{\alpha\beta}=0$; moreover,  the set
${\rm Sec}\,(\stackrel{k}{\oplus}TQ)= {\rm Sec}\,(\tkq)$ is  the  set
 $\mathfrak{X}^k(Q)$ of $k$-vectors fields on $Q$.
  \end{remark}

\subsubsection{The  Lagragian prolongation}\label{Lagprol}

For the description of the Lagrangian $k$-symplectic  formalism on lie algebroids we
consider the prolongation of a Lie algebroid $E$ over the fibration
$\widetilde{\tau}\colon\ke\to Q$, that is, (see Section \ref{prolong}),
\begin{equation}\label{te}\te=\{( {e}_{q}, {v}_{ \mathbf{{b}}_{q}})\in E\times T(\ke) /\;
\rho_E( {e}_{q})=T\widetilde{\tau}( {v}_{\mathbf{b}_{q}})\}\,,\end{equation}
where $ \mathbf{{b}}_{q}\in \stackrel{k}{\oplus} E_{q}$.
Taking into account the description of the prolongation
$\mathcal{T}^EP$ (see for instance,
\cite{CLMMM-2006,LMM-2005,Mart-2001}) on the particular case
$P=E\oplus\stackrel{k}{\ldots}\oplus E$ we obtain:
\begin{enumerate}
\item $\te\equiv E\times_{TQ}T(\ke)$ is a Lie algebroid over $\ke$ with projection
   $$\widetilde{\tau}_{\ke}\colon \te\equiv
E\times_{TQ}T(\ke)\longrightarrow \ke$$ and Lie algebroid structure
$(\lcf\cdot,\cdot\rcf^{\widetilde{\tau}},\rho^{\widetilde{\tau}}\,)$
where the anchor map
$$\rho^{\widetilde{\tau}}=E\times_{TQ}T(\ke)\colon\te\to T(\ke)$$ is
the canonical projection over the second factor.

We will refer to this particular Lie algebroid as the {\it Lagrangian prolongation}.

\item  If $(q^i,y^\alpha_A)$ denotes a local coordinate system of
   $\ke$ then the induced local coordinate system on
    $\te\equiv E\times_{TQ}T(\ke)$ is given by
     $$(q^i,y^\alpha_A,z^\alpha,w^\alpha_ A)_{1\leq i\leq n,\;1 \leq A\leq k,\; 1\leq \alpha\leq m}$$
where
\begin{equation}\label{local coord prol}\begin{array}{lcllcl}
 q^i({  e}_{q},{  v}_{ \mathbf{{b}}_{q}}) &=& q^i(q)
  \;,\quad &
y^\alpha_A({  e}_{q},{  v}_{ \mathbf{{b}}_{q}})&=&y^\alpha_A( \mathbf{{b}}_{q})\;,
\\\noalign{\medskip}
z^\alpha({  e}_{q},{  v}_{ \mathbf{{b}}_{q}})&=&y^\alpha(e_{q}) \;,\quad
& w^\alpha_A({  e}_{q},{  v}_{ \mathbf{{b}}_{q}})
&=&{  v}_{ \mathbf{{b}}_{q}}(y^\alpha_A)\;. \\
\end{array}
\end{equation}

\item The set
$\{\mathcal{X}_\alpha,\mathcal{V}^A_\alpha\}$ given by
\begin{equation}\label{base} \begin{array}{lccrcl}
\mathcal{X}_\alpha\colon &\ke&\to&\mathcal{T}^E(\ke) &\equiv &
E\times_{TQ}T(\ke)\\\noalign{\medskip}
 & \mathbf{{b}}_ {q}&\mapsto &
\mathcal{X}_\alpha( \mathbf{{b}}_ {q}) &=&
(e_\alpha( {q});\rho^i_\alpha( {q})\ds\frac{\partial
}{\partial q^i}\Big\vert_{ \mathbf{{b}}_ {q}})
\\\noalign{\bigskip}
\mathcal{V}^A_\alpha\,\colon &\ke&\to&\mathcal{T}^E(\ke)&\equiv &
E\times_{TQ}T(\ke)\\\noalign{\medskip}
 & \mathbf{{b}}_ {q}&\mapsto &
\mathcal{V}^A_\alpha( \mathbf{{b}}_ {q})
&=&({ {0}}_ {q};\ds\frac{\partial}{\partial
y^\alpha_A}\Big\vert_{ \mathbf{{b}}_ {q}})\,,
\end{array}\end{equation} is a local basis of $Sec(\te)$ the set of sections of $\widetilde{\tau}_{\ke}$. (See (\ref{base k-prol})).

\item The anchor map $\rho^{\widetilde{\tau}}\colon\te\to T(\ke)$ allows
 us to
associate a vector field to each section $\xi\colon\ke\to\te\equiv
E\times_{TQ}T(\ke)$ of $\widetilde{\tau}_{\ke}$  .

Locally, if $\xi$ writes as follows:
$$\xi=\xi^\alpha\mathcal{X}_\alpha+\xi^\alpha_A\mathcal{V}^A_\alpha\in Sec(\te)$$
then the associated vector field has the following local expression, see (\ref{rhoz}),
\begin{equation}\label{rholksim}
\rho^{\widetilde{\tau}}(\xi)=\rho^i_\alpha \xi^\alpha\derpar{}{q^i}
+ \xi^\alpha_A\derpar{}{y^\alpha_A}\in \vf(\ke)\,.\end{equation}

\item The Lie bracket of two sections  of $\widetilde{\tau}_{\ke}$
is characterized by the following expressions (see (\ref{lie brack k-prol})):
\begin{equation}\label{lie brack te}\begin{array}{lll}
\lcf\mathcal{X}_\alpha,\mathcal{X}_\beta\rcf^{\widetilde{\tau}}=
\mathcal{C}^\gamma_{\alpha\beta}\mathcal{X}_\gamma &
\lcf\mathcal{X}_\alpha,\mathcal{V}^B_\beta\rcf^{\widetilde{\tau}}=0
&
\lcf\mathcal{V}^A_\alpha,\mathcal{V}^B_\beta\rcf^{\widetilde{\tau}}=0\,,
\end{array}\end{equation}

\item If $\{\mathcal{X}^\alpha,
\mathcal{V}^\alpha_A\}$ is the dual basis of $\{\mathcal{X}_\alpha,
\mathcal{V}_\alpha^A\}$,  then the exterior differential is locally
given by, (see (\ref{dtp})),
\begin{equation}\label{difksim}
\begin{array}{lcl}
d^{\te}f&=&\rho^i_\alpha\derpar{f}{q^i}\mathcal{X}^\alpha +
\derpar{f}{y^\alpha_A}\mathcal{V}^\alpha_A\,,\quad \makebox{for all} \;
f\in \mathcal{C}^\infty(\ke)\\
d^{\te}\mathcal{X}^\gamma &=&
-\ds\frac{1}{2}\mathcal{C}^\gamma_{\alpha\beta}\mathcal{X}^\alpha\wedge\mathcal{X}^\beta,\quad
d^{\te}\mathcal{V}^\gamma_A =0\,.\end{array}\end{equation}
\end{enumerate}

\begin{remark}\label{remark equiv}{\rm
In the particular case $E= TQ$ the manifold $\mathcal{T}^E(\ke)$
turns into $T(\tkq)$.  In fact,  in this case we consider the prolongation of $TQ$
over $\taukq:\tkq\to Q$. Thus from (\ref{te}) we obtain
\begin{equation}\label{equiva}\begin{array}{cl}&\mathcal{T}^{TQ}(\stackrel{k}{\oplus}TQ)=\mathcal{T}^{TQ}(\tkq)
\\\noalign{\medskip} =&
\{({  u}_{q},{  v}_{{  \mathbf{w}}_{q}})\in TQ\times
T(\tkq)/ {  u}_{q}=T(\tau^k_Q)({  v}_{{
\mathbf{w}}_{q}})\}\\\noalign{\medskip} =&\{(T(\tau^k_Q)({  v}_{{
\mathbf{w}}_{q}}),{  v}_{{ \mathbf{ w}}_{q}})\in  TQ\times T(\tkq)/\; {  \mathbf{w}}_{q}\in
\tkq\}\\\noalign{\medskip}\equiv&\{{  v}_{{  \mathbf{w}}_{q}}\in T(\tkq)/\;
{  \mathbf{w}}_{q}\in \tkq\}\equiv T(\tkq)
\end{array}\end{equation} } \end{remark}

\subsubsection{The Liouville sections and the vertical
endomorphism.}\label{Lagform1}

On $\mathcal{T}^E(\ke)$ we are going to define two families of
canonical objects:   {\it  Liouville sections} and {\it vertical
endomorphism} which corresponds with   {\it  the Liouville vector
fields} and {\it the $k$-tangent structure} on $\tkq$, when we consider the particular case $E=TQ$. (See
\cite{Gu-1987,MRS-2004,RSV-2007}).

\paragraph{\bf Vertical $A$-lift.} (See, for instance \cite{CLMMM-2006}).
An element $({  e}_{q},{  v}_{{\mathbf{b}_ {q}}})$ of
$\mathcal{T}^E(\stackrel{k}{\oplus}E)\equiv E\times_{TQ}T(\ke)$ is
said to be {\it vertical}  if
\begin{equation}\label{vert}\widetilde{\tau}_1({  e}_{q},{  v}_{{{
\mathbf{b}}_ {q}}})={0}_ {q}\in E\,,\end{equation} where
$$\begin{array}{rrcl}
\widetilde{\tau}_1:&\mathcal{T}^E(\stackrel{k}{\oplus}E)\equiv E\times_{TQ}T(\ke)&\to& E,\\
&({  e}_{q},{  v}_{{{  \mathbf{b}}_ {q}}}) & \mapsto &\;
\widetilde{\tau}_1({  e}_{q},{  v}_{{{
\mathbf{b}}_ {q}}})={  e}_{q}\end{array}$$ is the projection on the first
factor $E$ of  $\mathcal{T}^E(\ke)$.

The above definition implies that the vertical elements of
$\mathcal{T}^E(\ke)$ are of the form
$$({0}_{q},{  v}_{{{  \mathbf{b}}_ {q}}})\in \mathcal{T}^E(\ke)\equiv
E\times_{TQ} T(\ke)$$ where ${  v}_{{{  \mathbf{b}}_ {q}}}\in
T(\stackrel{k}{\oplus}E)$ and ${  \mathbf{b}}_ {q}\in \ke$.

 Now, taking into account the definition (\ref{te}), which determines
 the elements of
$\mathcal{T}^E(\ke)$, the condition (\ref{vert}) means that
$${0}_{  q}=T_{{  \mathbf{b}}_{q}}\widetilde{\tau}\big({  v}_{{
\mathbf{b}}_{q}}\big)\,,$$ that is, the tangent vector ${  v}_{{{
\mathbf{b}}_ {q}}}$ is $\widetilde{\tau}$-vertical.

In a local coordinate system $(q^i,y^\alpha_A)$ on $\ke$, if
$({  e}_{  q},{  v}_{ {\mathbf{b}}_{q}})\in\te$ is vertical then ${  e}_{  q}={0}_{  q}$ and
$${  v}_{{{  \mathbf{b}}_ {q}}}=  u^\alpha_A \ds\frac{\partial }{\partial
y^\alpha_A}\Big\vert_{{{  \mathbf{b}}_ {q}}}\in T_{{{
\mathbf{b}}_ {q}}}( \stackrel{k}{\oplus}E)\,.$$

\begin{definition}\label{lvastke} For each
 $A=1,\ldots, k$ we call {\it the vertical $A^{th}$-lifting map} to the mapping
 \begin{equation}\label{a-levantamiento}
\begin{array}{rcl}
\xi^{V_A}:E\times_Q(\stackrel{k}{\oplus}E) & \longrightarrow &
\mathcal{T}^E(\stackrel{k}{\oplus}E)\equiv E\times_{TQ}T(\ke)
\\\noalign{\medskip}
 ({  e}_{q}, {\mathbf{b}}_ {q}) & \longmapsto &
\xi^{V_A} ({  e}_{q}, {\mathbf{b}}_ {q})=\left(0_ {q},({  e}_{q})^{V_A}_{{{ \mathbf{b}}_ {q}}}\right) \\
\end{array}
\end{equation} where ${  e}_{q}\in E,\;
 {\mathbf{b}}_{q}=({ {b}_1}_{q},\ldots, { {b}_k}_{q})\in \ke$ and the
 vector $\;({  e}_{q})^{V_A}_{{{
\mathbf{b}}_ {q}}}\in T_{{{  \mathbf{b}}_ {q}}} (\stackrel{k}{\oplus}E)$
is given by
\begin{equation}\label{vertical}
( {e}_{q})^{V_A}_{{{
\mathbf{b}}_ {q}}}f=\ds\frac{d}{ds}\Big\vert_{s=0}f(
 {b}_{1_ {q}},\ldots,  {b}_{A_ {q}}+s  {e}_{ {q}},\ldots,
 {b}_{k_ {q}})\;,\quad 1\leq A\leq k\;,
\end{equation} for an arbitrary function  $f\in \mathcal{C}^\infty(\stackrel{k}{\oplus}E)$.
\end{definition}

$\,$From (\ref{vertical}) we deduce that the local expression of
$({  e}_{q})^{V_A}_{{{  \mathbf{b}}_ {q}}}$ is the following:
\begin{equation}\label{localvert}
({  e}_{q})^{V_A}_{{{
\mathbf{b}}_ {q}}}=y^\alpha({  e}_{q})\ds\frac{\partial}{\partial
y^\alpha_A}\Big\vert_{{{ \mathbf{ b}}_ {q}}}\in T_{{{
\mathbf{b}}_ {q}}}( \stackrel{k}{\oplus}E)\;,\quad 1\leq A\leq k\;.
\end{equation}

On (\ref{localvert}) let us observe that the vector
$\;({  e}_{q})^{V_A}_{{{  \mathbf{b}}_ {q}}}\in T_{{{
\mathbf{b}}_ {q}}} (\stackrel{k}{\oplus}E)$ is
$\widetilde{\tau}$-vertical. Then
$\xi^{V_A}({  e}_{q}, {\mathbf{b}}_ {q})$ is a vertical element of
$\mathcal{T}^E(\ke)$.

 {\,}From (\ref{base}),  (\ref{a-levantamiento}) and (\ref{localvert})
  we obtain that $\xi^{V_A}$ has the following local expression:
\begin{equation}\label{localxia}
\begin{array}{lcl}\xi^{V_A}
({  e}_{q}, {\mathbf{b}}_ {q})&=&({0}_ {q},y^\alpha({  e}_{q})\ds\frac{\partial}{\partial
y^\alpha_A}\Big\vert_{{{  \mathbf{b}}_ {q}}}) =
y^\alpha({  e}_{q})\mathcal{V}^A_\alpha( {\mathbf{b}}_ {q})\,,\end{array}\quad
1\leq A\leq k\;.\end{equation}

\begin{remark}{\rm\
\begin{enumerate}
\item
In the standard case, that is, when $E=TQ$ and $\rho_{TQ}=id_{TQ}$, we
have that  given
  ${  e}_{q} \in T_{q}Q$ and ${  \mathbf{v}}_ {q}
  =({{  v}_1}_{q},\ldots,{{  v}_k}_{q})\in T^1_kQ$  one has
\[({  e}_{q} )^{V_A}_{{  \mathbf{v}}_ {q}}(f)=\ds\frac{d}{ds}\Big\vert_{s=0}f(
{  v}_{1_ {q}},\ldots, {  v}_{A_ {q}}+s {  e}_{q} ,\ldots,
{  v}_{k_ {q}})\;,\quad 1\leq A\leq k\;,\] that is, the $A^{th}$-
vertical lift to $\tkq$ of the tangent vector ${  e}_{{  q}}\in T_{  q} Q$ (see
for instance,   \cite{Gu-1987,MRS-2004,RSV-2007}).

\item In the particular case $k=1$ we obtain that
 $\xi^{V_1}\equiv\xi^V:E\times_Q E\to \mathcal{T}^EE$
  is the {\it vertical lifting map} introduced by E.
  Mart\'{\i}nez in
  \cite{Mart-2001}.
  \end{enumerate}}\end{remark}

 \paragraph{\bf The Liouville sections.}\label{Lagform1}The
 {\it $A^{th}$-Liouville section} $\widetilde{\Delta}_A$ is the section of
$\widetilde{\tau}_{\stackrel{k}{\oplus}E}:\mathcal{T}^E(\stackrel{k}{\oplus}E)\to
\stackrel{k}{\oplus}E$ given by
$$
\begin{array}{rcl}
\widetilde{\Delta}_A: \stackrel{k}{\oplus} E & \to &
\mathcal{T}^E(\stackrel{k}{\oplus}E)\equiv E\times_{TQ}
T(\ke)\\\noalign{\medskip} {{ \mathbf{ b}}_ {q}}&\mapsto
&\widetilde{\Delta}_A( {\mathbf{b}}_ {q})=\xi^{V_A}
(pr_A( {\mathbf{b}}_ {q}), {\mathbf{b}}_ {q})=\xi^{V_A}({{  b}_A}_{q}, {\mathbf{b}}_{q})
\end{array}\;,\qquad  1\leq A\leq k\;,
$$  where $ {\mathbf{b}}_{q}=({{  b}_1}_{q},\ldots,
{{  b}_k}_{q})\in \ke$ y $pr_A:\ke\to E$ is the canonical projection
over the $A^{th}$-copy of $E$ in $\ke$.

\, From the local expression (\ref{localxia}) of $\xi^{V_A}$ and
taking into account that
$$y^\alpha({{  b}_A}_{q})=y^\alpha_A({{  b}_1}_{q},\ldots,
{{  b}_k}_{q})=y^\alpha_ A( {\mathbf{b}}_{q})$$ we obtain that
$\widetilde{\Delta}_A$ has the following local expression
\begin{equation}\label{Liouville ksim}
\widetilde{\Delta}_A=\ds\sum_{\alpha=1}^m
y^\alpha_A\mathcal{V}^A_\alpha\;, \quad 1\leq A\leq k\;.
\end{equation}

\begin{remark} \label{liouvilleA ksim}{\rm
In the standard case,  we
obtain that each section $\widetilde{\Delta}_A$ turns in the
following vector field
$$\begin{array}{rccl}\Delta_A\colon& \tkq &\to&
T(\tkq)\\\noalign{\medskip}
 &  {\mathbf{v}}_{q}=({{  v}_1}_{q},\ldots, {{  v}_A}_{q}) & \mapsto &
 ({{  v}_A}_{q})^{V_A}_{ {\mathbf{v}}_{q}}\end{array}$$ that is with
 the $A^{th}$-canonical vector field on $\tkq$.}
\end{remark}

In the standard Lagrangian $k$-symplectic formalism, the canonical
vector fields   $\Delta_1,\ldots,$ $\Delta_k$ allow us to define the
energy function. In analogous way, as we will see in the sequel,  we will also
define de energy function using the Liouville sections
$\widetilde{\Delta}_1,\ldots, \widetilde{\Delta}_k$ in the Lie algebroid setting.

\paragraph{\bf Vertical endomorphism on $\mathcal{T}^E(\ke)$.} The
second important family of canonical geometric elements on $\te$ is
the family of vertical endomorphism $\widetilde{J}^1, \ldots,
\widetilde{J}^k$.

\begin{definition}\label{endvertA}
For each $A=1,\ldots, k$ we define the  {\it $A^{th}$-vertical
endomorphism} on $\mathcal{T}^E(\stackrel{k}{\oplus}E)\equiv
E\times_{TQ}T(\ke)$ as the mapping
\begin{equation}\label{jtildeA}\begin{array}{rccl}
\widetilde{J}^A:&\mathcal{T}^E(\stackrel{k}{\oplus}E) & \to &
\mathcal{T}^E(\stackrel{k}{\oplus}E)\\\noalign{\medskip}
&({  e}_{q},{  v}_{{{  \mathbf{b}}_ {q}}})&\mapsto
&\widetilde{J}^A({  e}_{q},{  v}_{{{  \mathbf{b}}_ {q}}})=\xi^{V_A}
({  e}_{q}, {\mathbf{b}}_ {q})\,,
\end{array}\end{equation} where
${  e}_{q}\in
E,\; {\mathbf{b}}_{q}=({{  b}_1}_{q},\ldots,{{  b}_k}_{q})\in\ke$ and
${  v}_{{{  \mathbf{b}}_ {q}}}\in T_{ {\mathbf{b}}_{q}}(\ke)$.
\end{definition}

\begin{lemma} Let $\{\mathcal{X}_\alpha,\;\mathcal{V}^A_\alpha\}$ be a local basis of ${\rm Sec}(\te)$ and let
$\{\mathcal{X}^\alpha,\;\mathcal{V}_A^\alpha\}$ be its dual basis.

Using this local basis, we obtain that the local expression of
$\widetilde{J}^A$ is given by the following expression:
\begin{equation}\label{localtildeJAksim}
\widetilde{J}^A=\ds\sum_{\alpha=1}^m\mathcal{V}^A_\alpha\otimes\mathcal{X}^\alpha\;,
\quad 1\leq A\leq k\;.
\end{equation}\end{lemma}

\proof From  (\ref{base}) and (\ref{localxia}) we obtain
$$
\begin{array}{lcl}
\widetilde{J}^A(\mathcal{X}_\alpha( {\mathbf{b}}_{q}))&=&\xi^{V_A}({  e}_\alpha(q), {\mathbf{b}}_q)
=y^\beta({  e}_\alpha(q))\mathcal{V}^A_\beta( {\mathbf{b}}_ {q})=\mathcal{V}^A_\alpha( {\mathbf{b}}_ {q})\;,
\\\noalign{\medskip}
\widetilde{J}^A(\mathcal{V}^B_\alpha( {\mathbf{b}}_{q}))&=&\xi^{V_A}({0}_{q}, {\mathbf{b}}_q)
={0}_{ {\mathbf{b}}_{q}}\,,
\end{array}
$$
 for each $A, B=1,\ldots, k,\, \alpha= 1\ldots, m$, where
$ {\mathbf{b}}_{q}\in\ke$ is an arbitrary element of $\ke$.\qed

\

\begin{remark}\
{\rm \begin{enumerate} \item When one writes the definition of
$\widetilde{J}^1,\ldots,\widetilde{J}^k$ in the particular case $E=
TQ$ and $\rho=id_{TQ}$ one obtains the canonical $k$-tangent
structure $J^1,\ldots, J^{\,k}$ on $\tkq$.

\item In the particular case $k=1$ we obtain the vertical
endomorphism $S$ on $\mathcal{T}^E(TQ)$, that is, on the
prolongation of the Lie algebroid  $E$ over $\tau_Q:TQ\to Q$. This
endormorphsim $S$ was defined by   E. Mart\'{\i}nez in
\cite{Mart-2001b}.\end{enumerate} }\end{remark}

\subsubsection{Second order partial differential
equations.}\label{Lagform2}

In the standard $k$-symplectic Lagrangian formalism one obtains the
solutions of the Euler-Lagrange equations as integral sections of
certain second order partial differential equations ({\sc sopde}) on
$\tkq$ .

In order to introduce the analogous objet in the $k$-symplectic
approach on Lie algebroids, now we are going to analyze the concept of {\sc sopde} in
the standard case. In this case a {\sc sopde} $\xi$ is a section of
the maps $$\begin{array}{rccl}\tau^{\,k}_{T^1_kQ}\colon &
T^1_k(T^1_kQ)&\to &T^1_kQ\\\noalign{\medskip}
 & ({{  v}_1}_{ {\mathbf{w}}_{q}},\ldots,{{  v}_k}_{ {\mathbf{w}}_{q}}) & \mapsto &{ {\mathbf{w}}_{q}}\end{array}$$ and
$$\begin{array}{rccl}T^1_k(\tau^{\,k}_Q)\colon &T^1_k(T^1_kQ)&\to&
T^1_kQ\\\noalign{\medskip}
&({{  v}_1}_{ {\mathbf{w}}_{q}},\ldots,{{  v}_k}_{ {\mathbf{w}}_{q}})&\mapsto
&(T_{ {\mathbf{w}}_{q}}(\taukq)({v_1}_{ {\mathbf{w}}_{q}}),\ldots,T_{ {\mathbf{w}}_{q}}(\taukq)({v_k}_{ {\mathbf{w}}_{q}}))
\end{array},$$ where $\taukq\colon \tkq\to Q$ denotes the canonical
projection of the tangent bundle of $k^1$-velocities.

 Returning to our
case, we know that: $(i)$ $\,\ke$ and $\te$ play the role of $\tkq$
and $T(\tkq)$, respectively; $(ii)$ $\; T^1_k(\tkq)$ is the Whitney
sum of $k$ copies of $T(\tkq)$. Then it is natural to think that the
Whitney sum of $k$ copies of $\te$, that is,
$$
(\mathcal{T}^E)^1_k(\ke)\colon=\mathcal{T}^E(\ke) \oplus \stackrel{k}{\ldots}\oplus\mathcal{T}^E(\ke)\,,
$$
 will play the role of
$T^1_k(\tkq)$.

Now, the natural question is: {\it what are the maps playing the role of
$\tau^k_{\tkq}$ and $T^1_k(\taukq)$, when one considers Lie
algebroids?}

Now we consider the following maps:
$$\begin{array}{rrcl}\widetilde{\tau}^k_{\ke}\colon &
(\mathcal{T}^E)^1_k(\ke)\equiv \mathcal{T}^E(\ke) \oplus \stackrel{k}{\ldots}\oplus\mathcal{T}^E(\ke)&\to &\ke\\\noalign{\medskip}
 & (({{  a}_1}_{q},{{  v}_1}_{ {\mathbf{b}}_{q}}),\ldots,({{  a}_k}_{q},{{  v}_k}_{ {\mathbf{b}}_{q}})) & \mapsto &{ {\mathbf{b}}_{q}}\end{array}$$ and
$$\begin{array}{rrcl}\widetilde{\tau}^k_1\colon &(\mathcal{T}^E)^1_k(\ke)\equiv \mathcal{T}^E(\ke) \oplus \stackrel{k}{\ldots}\oplus\mathcal{T}^E(\ke)&\to&
\ke\\\noalign{\medskip}
&(({{  a}_1}_{q},{{  v}_1}_{ {\mathbf{b}}_{q}}),\ldots,({{  a}_k}_{q},{{  v}_k}_{ {\mathbf{b}}_{q}}))&\mapsto
&({{  a}_1}_{q},\ldots,{{  a}_k}_{q})
\end{array}.$$

These two maps play the role of $\tau^k_{T^1_kQ}$ and $T^1_k(\tau^k_Q)$, respectively. In fact, when $E=TQ$ there exists a difeomorphism   between $T(T^1_kQ)$ and $\mathcal{T}^{TQ}(T^1_kQ)$ given by, (see remark \ref{remark equiv}),
$$\begin{array}{rcl}T(\tkq) &\equiv &
\mathcal{T}^{TQ}(\tkq)= (TQ)\times_{TQ} T(\tkq)\equiv T(\tkq)\\\noalign{\medskip}
{  v}_{ {\mathbf{w}}_{q}} & \equiv & (T_{{\rm
\mathbf{w}}_{q}}(\tau^k_Q)({  v}_{{
\mathbf{w}}_{q}}),{  v}_{{ \mathbf{w}}_{q}})\end{array}\,.$$

Thus \begin{itemize}
\item The map $$\widetilde{\tau}^{\,k}_{\stackrel{k}{\oplus}TQ}\colon
(\mathcal{T}^{TQ})^1_k(\tkq)\equiv T^1_k(\tkq) \to \tkq$$ corresponds to $\tau^{\,k}_{\tkq}\colon  T^1_k(\tkq)\to\tkq$ since
$$\begin{array}{ll}&\widetilde{\tau}_{\tkq}^{\,k}((T_{{
\mathbf{w}}_{q}}(\tau^k_Q)({{  v}_1}_{{
\mathbf{w}}_{q}}),{{  v}_1}_{{  \mathbf{w}}_{q}}),\ldots,(T_{{
\mathbf{w}}_{q}}(\tau^k_Q)({{  v}_k}_{{
\mathbf{w}}_{q}}),{{  v}_k}_{{  \mathbf{w}}_{q}}))={
\mathbf{w}}_{q}\\\noalign{\medskip}=&\tau^k_{\tkq}({{  v}_1}_{{
\mathbf{w}}_{q}},\ldots,{{  v}_k}_{{  \mathbf{w}}_{q}})\,.\end{array}
$$

\item The map
$$\widetilde{\tau}^{\,k}_1\colon
(\mathcal{T}^{TQ})^1_k(\tkq)\equiv T^1_k(\tkq) \to \tkq$$ identifies with $T^1_k(\taukq)\colon  T^1_k(\tkq)\to\tkq$ since $$\begin{array}{ll}
&\widetilde{\tau}_1^{\,k}((T_{{
\mathbf{w}}_{q}}(\tau^k_Q)({{  v}_1}_{{
\mathbf{w}}_{q}}),{{  v}_1}_{{  \mathbf{w}}_{q}}),\ldots,(T_{{
\mathbf{w}}_{q}}(\tau^k_Q)({{  v}_k}_{{
\mathbf{w}}_{q}}),{{  v}_k}_{{
\mathbf{w}}_{q}}))\\\noalign{\medskip}=&(T_{{
\mathbf{w}}_{q}}(\tau^k_Q)({{  v}_1}_{{
\mathbf{w}}_{q}}),\ldots,T_{{
\mathbf{w}}_{q}}(\tau^k_Q)({{  v}_k}_{{
\mathbf{w}}_{q}}))
=T^1_k(\taukq)({{  v}_1}_{{
\mathbf{w}}_{q}},\ldots,{{  v}_k}_{{  \mathbf{w}}_{q}})\,.\end{array}$$
\end{itemize}

\begin{remark}{\rm
For simplicity we denote by $( {\mathbf{a}}_{q}, {\mathbf{v}}_{ {\mathbf{b}}_{q}})$ an element
$$(({{  a}_1}_{q},{{  v}_1}_{ {\mathbf{b}}_q}),\ldots,
({{  a}_k}_{q},{{  v}_k}_{ {\mathbf{b}}_q}))$$ of $(\mathcal{T}^E)^1_k(\ke)\equiv \mathcal{T}^E(\ke) \oplus \stackrel{k}{\ldots}\oplus\mathcal{T}^E(\ke)$ where $ {\mathbf{a}}_{q}\colon =({{  a}_1}_{q},\ldots,{{  a}_k}_{q})\in \ke$ and
$ {\mathbf{v}}_{ {\mathbf{b}}_{q}}\colon =({{  v}_1}_{ {\mathbf{b}}_{q}},\ldots,{{  v}_k}_{ {\mathbf{b}}_{q}})\in
T^1_k(\ke)$.
}\end{remark}

Now we are in conditions to introduce the object which plays the role of a {\sc sopde} when we consider an arbitrary Lie algebroid $E$. This object is also called {\sc sopde}

\begin{definition} A {\rm second order partial differential equation} ({\sc sopde} for short) on $\ke$ is a map $\mathbf{\xi}\colon \ke\to (\mathcal{T}^E)^1_k(\ke)$ which is a section of $\widetilde{\tau}^k_{\ke}$ and $\widetilde{\tau}^k_1$.
\end{definition}
Since
$(\mathcal{T}^E)^1_k(\ke)$ is  the Whitney sum $\mathcal{T}^E(\ke) \oplus \stackrel{k}{\ldots}\oplus\mathcal{T}^E(\ke)$ of $k$ copies of $\mathcal{T}^E(\ke)$,
  we deduce that to give a section $\mathbf{\xi}$ of  $\widetilde{\tau}^k_{\ke}$ is equivalent to give
 a family of $k$ sections, $\xi_{1}, \dots, \xi_{k}$, of the Lagrangian prolongation $\mathcal{T}^E(\ke)$,   obtained by
projecting $\mathbf{\xi}$ on each factor.

Next, we are going to characterize a {\sc sopde}.
\begin{definition}
The set \begin{equation}\label{adm}
\begin{array}{lcl}
Adm(E)&=&\{( {\mathbf{a}}_{q}, {\mathbf{v}}_{{  \mathbf{b}}_ {q}})\in
(\mathcal{T}^E)^1_k(\stackrel{k}{\oplus}E)\;|\;\widetilde{\tau}_1^{\,k}( {\mathbf{a}}_{q}, {\mathbf{v}}_{{
\mathbf{b}}_ {q}})
=\widetilde{\tau}^{\,k}_{\stackrel{k}{\oplus}E}( {\mathbf{a}}_{q}, {\mathbf{v}}_{{
\mathbf{b}}_ {q}})\}\\\noalign{\medskip}
&=&\{( {\mathbf{a}}_{q}, {\mathbf{v}}_{{  \mathbf{b}}_ {q}})\in
(\mathcal{T}^E)^1_k(\stackrel{k}{\oplus}E)\;|\; {\mathbf{a}}_{q}={
\mathbf{b}}_ {q}\}\;.\end{array}
\end{equation}
is called the {\rm set of admissible points}.
\end{definition}

\begin{proposition}\label{sopdechar}
Let $\mathbf{\xi}=(\xi_1,\ldots, \xi_k):\stackrel{k}{\oplus}E\to
(\mathcal{T}^E)^1_k(\stackrel{k}{\oplus}E)$ be a section of
$\widetilde{\tau}^k_{\ke}$. The following statements
are equivalent:

\begin{enumerate}
   \item $\mathbf{\xi}$ takes values in $Adm(E)$ .
   \item $\mathbf{\xi}$ is a {\sc sopde}, that is, $\widetilde{\tau}^k_1\circ \mathbf{\xi}=id_{\stackrel{k}{\oplus}E}$ .
   \item $\widetilde{J}^A(\xi_A)=\widetilde{\Delta}_A$
for all $A=1,\ldots, k$.
\end{enumerate}

\end{proposition}

\proof {}From (\ref{adm}) it is easy to prove that (i) and (ii) are
equivalent. The equivalence between (i) and (iii) is a direct consequence
of the definitions of $\widetilde{J}^A,\,\widetilde{\Delta}_A$  and
$\xi^{V_A}$. \qed

Using $(iii)$ in Proposition \ref{sopdechar} one easily  deduce that the
local expression of a {\sc sopde} $\mathbf{\xi}=(\xi_1,\ldots, \xi_k)$ is the following
$$
\xi_A=y^\alpha_A\mathcal{X}_\alpha+
(\xi_A)^\alpha_B\mathcal{V}^B_\alpha
$$ where $(\xi_A)^\alpha_B\in \mathcal{C}^\infty(\ke)$.

\begin{proposition}
Let $\mathbf{\xi}=(\xi_1,\ldots, \xi_k):\stackrel{k}{\oplus}E\to
(\mathcal{T}^E)^1_k(\stackrel{k}{\oplus}E)$ be a section of
$\widetilde{\tau}^k_{\ke}$. Then $$(\rho^{\widetilde{\tau}}(\xi_1),\ldots,\rho^{\widetilde{\tau}}(\xi_k))\colon \ke\to
T^1_k(\ke)$$ is a $k$-vector field on $\ke$. Let us remember that $$\rho^{\widetilde{\tau}}\colon \te\equiv
E\times_{TQ}T(\ke)\to T(\ke)$$ denote the anchor map of the Lie algebroid $\te$.
\end{proposition}
\proof It is a direct consequence of $(vi)$ in Section \ref{Lagprol}.\qed

In local coordinate we obtain
\begin{equation}\label{sopde asso}
\rho^{\widetilde{\tau}}(\xi_A)=\rho^i_\alpha
y^\alpha_A\ds\frac{\partial}{\partial
q^i}+(\xi_A)^\alpha_B\ds\frac{\partial}{\partial y^\alpha_B}\in \mathfrak{X}(\ke)\;.
\end{equation}
\begin{definition}
A map $$\eta:\rk\to \stackrel{k}{\oplus}E$$ es called an  {\rm
integral section}  of the  {\sc sopde} $\xi$, if $\eta$ is an integral section of the $k$-vector field $(\rho^{\widetilde{\tau}}(\xi_1)
,\ldots,\rho^{\widetilde{\tau}}(\xi_k))$, associated to $\xi$, that is,
\begin{equation}\label{int sect}
(\rho^{\widetilde{\tau}}(\xi_A))(\eta( {\mathbf{t}}))=
\eta_*( {\mathbf{t}})\left(\ds\frac{\partial}{\partial
t^A}\Big\vert_{ \mathbf{t}}\right)\;,\quad 1\leq A\leq k\;,
\end{equation}\end{definition}

If $\eta$ is written locally as
$\eta(\mathbf{t})=(\eta^i(\mathbf{t}),\eta^\alpha_A(\mathbf{t}))$, then from (\ref{sopde asso})
 we deduce that (\ref{int sect}) is locally equivalent
to the identities,
\begin{equation}\label{integral sect}
\ds\frac{\partial \eta^i}{\partial t^A}\Big\vert_{\mathbf{t}}=
\eta^\alpha_A(\mathbf{t})\rho^i_\alpha(\widetilde{\tau}(\eta(\mathbf{t}))\;,\quad
\ds\frac{\partial \eta^\beta_B}{\partial
t^A}\Big\vert_{\mathbf{t}}=(\xi_A)^\beta_B(\eta(\mathbf{t}))\;,
\end{equation}where $\widetilde{\tau}:\stackrel{k}{\oplus}E\to Q$ is the
canonical projection.
\subsubsection{Lagrangian formalism.}\label{Lagform3}

Let $L:\stackrel{k}{\oplus} E\to \r$ be a function which we will
call Lagrangian function.

In this section, we will develop a intrisic and global geometric framework, which allows
us to write the Euler-Lagrange equations  on a
Lie algebroid, associated with the Lagrangian function $L$. In first place we are going to introduce some geometric elements associated with a Lagrangian $L$.

\paragraph{\bf Poincar\'{e}-Cartan sections.}
We now introduce {\it the Poincar\'{e}-Cartan $1$-sections }
$$
\begin{array}{rcc}
\Theta_L^A: \stackrel{k}{\oplus}E & \longrightarrow &
(\mathcal{T}^E(\stackrel{k}{\oplus}E))^{\;*}
\\\noalign{\medskip}
{ {\mathbf{b}}_ {q}} & \longmapsto & \Theta_L^A( {\mathbf{b}}_ {q})\end{array}
$$
 where $\Theta_L^A( {\mathbf{b}}_ {q})$ is defined by
$$\begin{array}{rlcl}
\Theta_L^A( {\mathbf{b}}_ {q}): & (\mathcal{T}^E(\stackrel{k}{\oplus}E)
)_{{ {\mathbf{b}}_ {q}}}& \longrightarrow & \r
\\\noalign{\medskip}
   & Z_{{ {\mathbf{b}}_ {q}}}=({  e}_{q},{  v}_{ {\mathbf{b}}_{q}}) & \longmapsto &
(\Theta_L^A)_{{{  \mathbf{b}}_ {q}}}(Z_{{{  \mathbf{b}}_ {q}}})=(d^{\mathcal{T}^E(\ke)}L)_{{{  \mathbf{b}}_ {q}}}((\widetilde{J}^A)_{{{  \mathbf{b}}_ {q}}}(Z_{{  \mathbf{b}}_ {q}}))
\end{array}\;.
$$
Using the expression (\ref{difksim}) of $d^{\te}f$ with $f=L$
we obtain:
\begin{equation}\label{theta al 2}
(\Theta_L^A)({{  \mathbf{b}}_ {q}})Z_{{{  \mathbf{b}}_ {q}}}=(d^{\mathcal{T}^E(\ke)}
L)_{{{  \mathbf{b}}_ {q}}}\Big((\widetilde{J}^A)_{{{
\mathbf{b}}_ {q}}}Z_{{{  \mathbf{b}}_ {q}}}\Big)
=\Big(\rho^{\widetilde{\tau}}\big((\widetilde{J}^A)_{{{
\mathbf{b}}_ {q}}}Z_{{{  \mathbf{b}}_ {q}}}\big)\Big)L\;,
\end{equation}
where ${{  \mathbf{b}}_ {q}}\in
\stackrel{k}{\oplus}E,\;Z_{{{  \mathbf{b}}_ {q}}}\in
[\mathcal{T}^E(\stackrel{k}{\oplus}E)]_{{{  \mathbf{b}}_ {q}}}$ y
$\rho^{\widetilde{\tau}}((\widetilde{J}^A)_{{{
\mathbf{b}}_ {q}}}Z_{{{  \mathbf{b}}_ {q}}})\in T_{{
\mathbf{b}}_ {q}}(\ke)$.

The Poincar\'{e}-Cartan $2$-sections
$$\Omega_L^A:\stackrel{k}{\oplus}E \to
(\mathcal{T}^E(\stackrel{k}{\oplus}E))^{\;*}\wedge(\mathcal{T}^E(\stackrel{k}{\oplus}E))^{\;*},\;1\leq
A\leq k$$ are defined as follows:
$$
\Omega_L^A\colon  =-\d^{\mathcal{T}^E(\ke)}\Theta_L^A\;,\quad 1\leq A\leq k\;,
$$
that is,
\begin{equation}\label{OmegaLA}\begin{array}{lcl}\Omega_L^A(\xi_1,\xi_2)&=& -
\d\Theta_L^A(\xi_1,\xi_2)\\\noalign{\medskip} &=&
[\rho^{\widetilde{\tau}}(\xi_2)](\Theta_L^A(\xi_1)) -
[\rho^{\widetilde{\tau}}(\xi_1)](\Theta_L^A(\xi_2)) +
\Theta_L^A(\lcf\xi_1,\xi_2\rcf^{\widetilde{\tau}})\,,\end{array}\end{equation}
where $\xi_1,\xi_2\in Sec(\te)$ and
$(\rho^{\widetilde{\tau}},\lcf\cdot,\cdot\rcf^{\widetilde{\tau}})$
denotes the Lie algebroid structure of $\te$ defined in section \ref{Lagprol}.

Next, we will establish the local expressions of
  $\Theta_L^A$ and $\Omega_L^A$.

Consider $\{\mathcal{X}_\alpha,\;\mathcal{V}^B_\alpha\}$ a local basis of sections of
${\rm Sec}(\te)$ and
$\{\mathcal{X}^\alpha,\;\mathcal{V}_B^\alpha\}$ its dual basis. Then from
(\ref{rholksim}), (\ref{localtildeJAksim}) y (\ref{theta al 2})
we obtain
\begin{equation}\label{local theta}
\Theta_L^A=\ds\frac{\partial L}{\partial y^\alpha
_A}\mathcal{X}^\alpha \;,\qquad 1\leq A\leq k\;.
\end{equation}

{} From de local expressions (\ref{base}), (\ref{rholksim}),
(\ref{lie brack te}),  (\ref{OmegaLA}) and (\ref{local theta})
we have for each $A=1, \ldots, k$,
{\small\begin{equation}\label{local omega} \Omega_L^A =
\ds\frac{1}{2} \left(\rho^i_\beta \ds\frac{\partial ^2 L}{\partial
q^i\partial y^\alpha_A} - \rho^i_\alpha \ds\frac{\partial ^2
L}{\partial q^i\partial y^\beta_A}+
\mathcal{C}^\gamma_{\alpha\beta}\ds\frac{\partial L}{\partial
y^\gamma_A}\right) \mathcal{X}^\alpha \wedge \mathcal{X}^\beta +
\ds\frac{\partial ^2 L}{\partial y^\beta_ B\partial y^\alpha_A}\,
\mathcal{X}^\alpha \wedge \mathcal{V}_B^\beta\;.
\end{equation}}

\begin{remark}{\rm\
\begin{enumerate}
\item En the particular case $k=1$ we obtain the Poincar\'{e}-Cartan forms of the Lagrangian Mechanics on Lie algebroids. See, for instance
\cite{CLMMM-2006,Mart-2001b}.

\item When $E= TQ$ and $\rho_{TQ}=id_{TQ}$,
then $$\Omega_L^A(X,Y)=\omega_L^A(X,Y)\,,\quad  1\leq A\leq k$$
where $X,Y$ are two vector fields on $\tkq$ and $\omega_L^1,\ldots,
\omega_L ^{\,k}$ denote the Lagrangian $2$-forms of the standard $k$-symplectic formalism defined by
$\omega_L^A=-d(dL\circ J^A)$, being $d$ the usual differential.
\end{enumerate}
}\end{remark}

\paragraph{\bf The energy function.}
The {\it energy function} $E_L:\stackrel{k}{\oplus}E \to \r$ defined
by the Lagrangian $L$ is
$$
E_L=\ds\sum_{A=1}^k\rho^{\widetilde{\tau}}(\Delta_A)L-L\;,
$$
and from (\ref{rholksim}) and (\ref{Liouville ksim}) one deduce that
$E_L$ is locally given by
\begin{equation}\label{local ener}
E_L=\ds\sum_{A=1}^k y^\alpha_A\ds\frac{\partial L}{\partial
y^\alpha_A}- L\;.
\end{equation}

\paragraph{\bf Morphisms.}
For studying the concept of  Euler-Lagrange equations and their solutions on Lie algebroids,
we need to show a new point of view of the solutions for the standard
Euler-Lagrange equations, which allows us to think a solution as a particular set of
  Lie algebroid morphisms.

In the standard Lagrangian $k$-symplectic  formalism, a solution of the Euler-Lagrange
equation is a field $\phi:\rk\to Q$ such that its first prolongation
$\phi^{(1)}:\rk\to T^1_kQ$ satisfies the Euler-Lagrange field
equations, that is,
$$
\displaystyle \sum_{A=1}^k\ds\frac{\partial}{\partial
t^A}\Big\vert_{ {\mathbf{t}}} \left(\frac{\displaystyle\partial
L}{\displaystyle
\partial v^i_A}\Big\vert_{\phi^{(1)}( {\mathbf{t}})} \right)= \frac{\displaystyle \partial
L}{\displaystyle
\partial q^i}\Big\vert_{\phi^{(1)}( {\mathbf{t}})}\,.
$$

Let us observe that the map $\phi$ naturally induces  the following Lie algebroid morphism
\[\xymatrix
{T\rk  \ar[r]^-{T\phi}\ar[d]_{\tau_{\rk}} & TQ\ar[d]^-{\tau_Q}\\
\rk\ar[r]_-{\phi} & Q}\]

If we consider the canonical basis of section of $\tau_{\rk}$,
$\;\left\{\ds\frac{\partial}{\partial t^1},\ldots,
\ds\frac{\partial}{\partial t^k}\right\}$, then the  first
prolongation $\phi^{(1)}$ of $\phi$, can be written as follows:
\[\phi^{(1)}(\mathbf{t})=(T_\mathbf{t}\phi(\ds\frac{\partial}{\partial
t^1}\Big\vert_{\mathbf{t}}),\ldots,T_\mathbf{t}\phi(\ds\frac{\partial}{\partial
t^k}\Big\vert_{\mathbf{t}}))\,.\]

Returning to the case of Lie algebroids, the analog of a field
solution of the Euler-Lagrange equations  is now a
 Lie algebroid morphism $\Phi=(\overline{\Phi},\underline{\Phi})$ between $T\rk$ and $E$
\[\xymatrix
{T\rk  \ar[r]^-{\overline{\Phi}}\ar[d]_{\tau_{\rk}} & E\ar[d]^-{\tau}\\
\rk\ar[r]_-{\underline{\Phi}} & Q}\]

Taking a local basis $\{e_A\}_{A=1}^k$ of local sections of $T\rk$,
one can define a map $\widetilde{\Phi}:\rk\to\stackrel{k}{\oplus} E$
associated to $\Phi$ and given by
\[\begin{array}{rcl}
\widetilde{\Phi}:\rk &\to & \stackrel{k}{\oplus}E\equiv
E\oplus\stackrel{k}{\ldots}\oplus E\\
\mathbf{t} &\to &(\overline{\Phi}(e_1(\mathbf{t})),\ldots, \overline{\Phi}(e_k(\mathbf{t})))\;.
\end{array}\]

Let $(t^A)$ and $(q^i)$ be a local coordinate system on $\rk$ and
$Q$, respectively. Let $\{e_A\}$ be a local basis of sections of
$\tau_{\rk}$ and $\{e_\alpha\}$ be a local basis of sections of $E$,
we denote by $\{e^A\}$ and $\{e^\alpha\}$ the dual basis. Then
$\Phi$ is determined by the relations
$\underline{\Phi}(\mathbf{t})=(\phi^i(\mathbf{t}))$ and $\Phi^* e^\alpha=\phi^\alpha_A
e^A$ for certain local functions $\phi^i$ and $\phi^\alpha_A$ on
$\rk$. Thus, the associated map $\widetilde{\Phi}$ is locally given
by $\widetilde{\Phi}(\mathbf{t})=(\phi^i(\mathbf{t}),\phi^\alpha_A(\mathbf{t}))$.

In this case, the conditions of Lie algebroid morphism (\ref{morp cond}) are written as\begin{equation}\label{morpcond}
  \rho^i_
 \alpha\phi^\alpha_A = \ds\frac{\partial \phi^i}{\partial t^A} \quad , \quad
0=\ds\frac{\partial \phi^\alpha_A}{\partial t^B} -
\ds\frac{\partial \phi^\alpha_B}{\partial t^A}
+\mathcal{C}^\alpha_{\beta\gamma}\phi^\beta_B\phi^\alpha_A
\,.\end{equation}

\begin{remark} In the standard case where $E=TQ$
the above morphism conditions reduce to
\[\phi^i_A=\ds\frac{\partial \phi^i}{\partial t^A} \quad
\makebox{and}\quad  \ds\frac{\partial \phi^i_A}{\partial
t^B}=\ds\frac{\partial \phi^i_B}{\partial t^A}\,.\] Therefore, in
the standard case, by considering morphisms we are just considering
the first-order prolongation of the fields $\phi:\rk\to Q$.\end{remark}

\paragraph{\bf \bf The Euler-Lagrange equations.}

Consider a given regular Lagrangian function $L\colon \ke\to \R.$ The field equations are obtained as follows:

We look for the solutions $\mathbf{\xi}=(\xi_1,\ldots,\xi_k)$ of the equation
\begin{equation}\label{ec ge EL}
 \ds\sum_{A=1}^k\imath_{\xi_A}\Omega_L^A=\d^{\mathcal{T}^E(\ke)} E_L\;.
\end{equation}

Notice that each $\xi_A$ is a section of the Lagrangian prolongation $\te$ and thus, $\mathbf{\xi}$ is a section of $(\mathcal{T}^E)^1_k(\ke)=\te\oplus\stackrel{k}{\ldots}\oplus\te\to \ke$.

Using a local coordinate system $(q^i, y^\alpha_A)$ on $\ke$ an a local basis $\{e_\alpha\}$ of ${\rm Sec}(E)$, each $\xi_A$ is locally given by
$$\xi_A= \xi_A^\alpha \mathcal{X}_\alpha +
(\xi_A)^\alpha_C \mathcal{V}^C_\alpha\,. $$

Then, using this local expression and from (\ref{difksim}), (\ref{local omega}) and (\ref{local ener}) we obtain that the equation (\ref{ec ge EL}) is locally expressed as follows:
 {\small$$\begin{array}{rcl}
\xi^\beta_A\Big(\rho^i_\alpha\,\derpars{L}{q^i}{y^\beta_A}
-\rho^i_\beta\,\derpars{L}{q^i}{y^\alpha_A} + {\mathcal
C}^\gamma_{\beta\alpha}\derpar{L}{y^\gamma_A}\Big)\,- (\xi_A)^\beta_
B\derpars{L}{y^\beta_ B}{y^\alpha_A}\, &=&
\rho^i_\alpha\Big(y^\beta_A\derpars{L}{q^i}{y^\beta_A}
-\derpar{L}{q^i}\Big),\\\noalign{\medskip}\xi^\alpha_A
\derpars{L}{y^\beta_ B}{y^\alpha_A} &=& y^\alpha_
A\derpars{L}{y^\beta_ B}{y^\alpha_A}\,.
\end{array}$$}

Since $L$ is regular, that is the matrix $(\frac{\partial^2L}{\partial y^\alpha_A\partial y^\beta_B})$ is regular, the above equations can be written as follows
\begin{equation}\label{cond sopde ksim}
\begin{array}{rcl}
y^\beta_A\rho^i_\beta \ds\frac{\partial ^2 L}{\partial q^i\partial
y^\alpha_A} + (\xi_A)^\beta_B\ds\frac{\partial^2 L}{\partial
y^\alpha_A\partial  y^\beta_B} &=& \rho^i_\alpha \ds\frac{\partial
L}{\partial q^i} +
y^\beta_A\mathcal{C}^\gamma_{\beta\alpha}\ds\frac{\partial
L}{\partial y^\gamma_A}\;,
\\\noalign{\medskip}
\xi^\alpha_A &=& y^\alpha_A\,.
\end{array}
\end{equation}Therefore $\mathbf{\xi}$ is a {\sc sopde}.

Let $\widetilde{\Phi}\colon \R^k\to \ke$ the associated map to a  Lie algebroid morphism $\Phi\colon T\R^k\to E$.

 If
$\widetilde{\Phi}( {\mathbf{t}})=(\phi^i( {\mathbf{t}}),
\phi^\alpha_A( \mathbf{{t}}))$ is an integral section of the  {\sc sopde}
$\mathbf{\xi}$ solution of (\ref{ec ge EL}) then from the condition (\ref{integral sect}) of integral section and
 the equations  (\ref{cond sopde ksim}) we obtain
\begin{eqnarray*}
 \ds\frac{\partial \phi^i}{\partial t^A}\Big\vert_{ \mathbf{{t}}} \ds\frac{\partial ^2 L}{\partial q^i\partial
y^\alpha_A}\Big\vert_{\widetilde{\Phi}( \mathbf{{t}})} +
\ds\frac{\partial \phi^\beta_B}{\partial
t^A}\Big\vert_{ \mathbf{{t}}}\ds\frac{\partial^2 L}{\partial
y^\alpha_A\partial
y^\beta_B}\Big\vert_{\widetilde{\Phi}( \mathbf{{t}})} &=& \rho^i_\alpha
\ds\frac{\partial L}{\partial
q^i}\Big\vert_{\widetilde{\Phi}( \mathbf{{t}})} +
\phi^\beta_A\mathcal{C}^\gamma_{\beta\alpha}\ds\frac{\partial
L}{\partial y^\gamma_A}\Big\vert_{\widetilde{\Phi}( \mathbf{{t}})}\;,
\\\noalign{\medskip}
  \ds\frac{\partial \phi^i}{\partial t^A}\Big\vert_{ \mathbf{{t}}} &=&\rho^i_\alpha
  \phi^\alpha_A( \mathbf{{t}})\;, \\
  0&=&\ds\frac{\partial \phi^\alpha_A}{\partial t^B}\Big\vert_{ \mathbf{t}} -
\ds\frac{\partial \phi^\alpha_B}{\partial t^A}\Big\vert_{ \mathbf{t}}
+\mathcal{C}^\alpha_{\beta\gamma}\phi^\beta_B( \mathbf{t})\phi^\gamma_A( \mathbf{t})
\end{eqnarray*} where the later equation is a consequence of the morphism condition
 (\ref{morpcond}). The above equations can be written as follows:
   \begin{equation} \label{eq E-L ksim}
\begin{array}{rcl}
\ds\sum_{A=1}^{\,k}\ds\frac{\partial}{\partial t^A}\left(
\ds\frac{\partial  L}{\partial
y^\alpha_A}\Big\vert_{\widetilde{\Phi}( {\mathbf{t}})}\right)  &=&
\rho^i_\alpha \ds\frac{\partial L}{\partial
q^i}\Big\vert_{\widetilde{\Phi}( {\mathbf{t}})} +
\phi^\beta_C\mathcal{C}^\gamma_{\beta\alpha}\ds\frac{\partial
L}{\partial
y^\gamma_C}\Big\vert_{\widetilde{\Phi}( {\mathbf{t}})}\\\noalign{\medskip}
  \ds\frac{\partial \phi^i}{\partial t^A}\Big\vert_{ {\mathbf{t}}} &=&
\rho^i_\alpha  \phi^\alpha_A( \mathbf{{t}})\;,
  \\\noalign{\medskip}
  0&=&\ds\frac{\partial \phi^\alpha_A}{\partial t^B}\Big\vert_{ \mathbf{t}} -
\ds\frac{\partial \phi^\alpha_B}{\partial t^A}\Big\vert_{ \mathbf{t}}
+\mathcal{C}^\alpha_{\beta\gamma}\phi^\beta_B( \mathbf{t})\phi^\gamma_A( \mathbf{t})\;.
\end{array}\end{equation}

 Notice that if $E$ is the standard Lie algebroid $TQ$
then the above equations are the classical Euler-Lagrange equations
in field theories for the Lagrangian $L: T^1_kQ\to \r$. Thus, in the sequel, (\ref{eq E-L ksim}) will be called the {\it Euler-Lagrange equations for field theories on Lie
algebroids}.

\begin{remark}{\rm \

\begin{enumerate}

    \item The equations (\ref{eq E-L ksim}) are obtained by E. Martinez from a variational approach in the multisymplectic framework, see \cite{Mart-2005}.

   \item  If one rewrite the above equations in the particular case, $k=1$,
one obtain the Euler-Lagrange equations on Lie algebroids given by  Weinstein  in
\cite{Weins-1996}.

\item When $E=TQ$, the equations (\ref{eq E-L ksim}) coincides with the
Euler- Lagrange equations of the G\"{u}nther formalism,
\cite{MRS-2004}.
    \end{enumerate} }\end{remark}

The  results of this section  can be summarized in the following

\begin{theorem}\label{algeform}
Let $L:\rk\to\stackrel{k}{\oplus}E$ be a regular Lagrangian and
$\xi_1,\ldots, \xi_k$ be $k$ sections of $\widetilde{\tau}_{\ke}\colon \te\to \ke$  such that
\[
\ds\sum_{A=1}^k\imath_{\xi_A}\Omega_L^A=\d^{\mathcal{T}^E(\ke)} E_L\;.\]
Then:
\begin{enumerate}
\item $\mathbf{\xi}=(\xi_1,\ldots, \xi_k)$ is a {\sc sopde}.
\item Let $\widetilde{\Phi}:\rk\to \stackrel{k}{\oplus}E\,$ be  the
map associated with a  Lie algebroid morphism between $T\rk$ and
$E$. If $\widetilde{\Phi}$ is an integral section of $\mathbf{\xi}$, then it
is a solution of the {\it  Euler-Lagrange   equations for field theories on Lie algebroids
(\ref{eq E-L ksim})}.
\end{enumerate}
\end{theorem}

\begin{remark}If we rewrite this section in the particular case $k=1$, we reobtain the Lagrangian Mechanics on a Lie algebroid.
(See section 3.1 in \cite{CLMMM-2006} or section 2.2 in
\cite{LMM-2005}).
\end{remark}

As a final remark in this subsection, it is interesting to point out
that the standard Lagrangian $k$-symplectic formalism   is a
particular case of the Lagrangian formalism on Lie algebroids,
 when $E=TQ$,  the anchor map $\rho_{TQ}$ is the identity
on $TQ$ and the structure constants are $\mathcal{C}_{\alpha
\beta}^\gamma=0$.

 In this case we have:
 \begin{itemize}
\item The manifold $\ke$ identifies with
$T^1_kQ$, $\mathcal{T}^{TQ}(\tkq)$ with $T(\tkq)$ and
$(\mathcal{T}^{TQ})^1_k(T^1_kQ)$ with $T^1_k(T^1_kQ)$.

\item The energy function $E_L:T^1_kQ\to \r$ is given by
 $E_L=\ds\sum_{A=1}^k\Delta_A(L)-L$ where
 $\Delta_A$ are the canonical vector field on $T^1_kQ$. We have explained how to obtain this vector fields in Remark \ref{liouvilleA ksim} .

\item A section $\mathbf{\xi}:\stackrel{k}{\oplus}E \to
(\mathcal{T}^E)^1_k(\stackrel{k}{\oplus}E)$ correspond to a $k$-vector field
 $\mathbf{\xi}=(\xi_1,\ldots,\xi_k)$ on $T^1_kQ$, that is, $\xi$ is a section of $\tau^{\,k}_{T^1_kQ}:T^1_k(T^1_kQ) \to
T^1_kQ$.

\item A {\sc  sopde} $\mathbf{\xi}$ is a $k$-vector field on $T^1_kQ$ which is a section of
$T^1_k(\tau^{\,k}_Q):T^1_k(T^1_kQ) \to T^1_kQ$.

\item Let $f$  be a function on $T^1_kQ$ then
$$\d^{\mathcal{T}^{TQ}(\tkq)}f(Y)=df(Y)\,,$$
where $df$ denotes the standard differential and $Y$ is a vector field on $T^1_kQ$.

\item It is satisfies that $$\Omega_L^A(X,Y)=\omega_L^A(X,Y),\qquad A=1,\ldots,k$$
where $\omega_L^A,\, A=1,\ldots, k$ are the Lagrangian $2$-forms  of the standard $k$-symplectic formalism given by
$\omega_L^A=-d(dL\circ J^A)$.

\item Thus, in the standard  $k$-symplectic Lagrangian formalism, the equation   (\ref{ec ge EL}) can be written as follow:
$$\ds\sum_{A=1}^{\,k}\imath_{\xi_A}\omega_L^A=dE_L\,,$$ that is, this equation is the geometric Euler-Lagrange equations
in the standard $k$-symplectic Lagrangian formalism.

\item In the standard case a map $\phi:\rk\to Q$ induces a
 Lie algebroid morphism $(T\phi,\phi)$ between $T\rk$ and $TQ$. In this
case, the associated map $\widetilde{\Phi}$ of this morphism is
 the first prolongation $\phi^{(1)}$ of $\phi$ given by
$$\widetilde{\Phi}( {\mathbf{t}})=(T\phi(\ds\frac{\partial}{\partial
t^1}\Big\vert_{ {\mathbf{t}}}),\ldots, T\phi(\ds\frac{\partial}{\partial
t^k}\Big\vert_{ {\mathbf{t}}}))\,.$$ Let us observe that
$\widetilde{\Phi}=\phi^{(1)}$ (see \ref{1prolong}).
\end{itemize}

Thus, from the Theorem \ref{algeform} and the above remarks, we
deduce the following corollary which summarizes the standard Lagrangian
$k$-symplectic formalism, see \cite{Gu-1987,MRS-2004,RSV-2007}.

\begin{corollary}Let $L:T^1_kQ\to \r$ be a regular Lagrangian and
$\mathbf{\xi}=(\xi_1,\ldots, \xi_k)$ a $k$-vector field on $T^1_kQ$ such that
$$\ds\sum_{A=1}^ki_{\xi_A}\omega_L^A=dE_L\,.$$
Then:
\begin{enumerate}
\item $\mathbf{\xi}$ is a {\sc sopde}
\item  If $\widetilde{\Phi}\equiv\phi^{(1)}$ is an integral section of
the $k$-vector field $\xi$, then it is a solution of the
Euler-Lagrange field equations in the standard Lagrangian
k-symplectic field theories given by
$$
\displaystyle \sum_{A=1}^k\ds\frac{\partial}{\partial t^A}\Big\vert_{\mathbf{t}}
\left(\frac{\displaystyle\partial L}{\displaystyle
\partial v^i_A}\Big\vert_{\widetilde{\Phi}( {\mathbf{t}})} \right)= \frac{\displaystyle \partial
L}{\displaystyle
\partial q^i}\Big\vert_{\widetilde{\Phi}( {\mathbf{t}})} \quad , \quad
v^i_A(\widetilde{\Phi}( {\mathbf{t}}))=  \frac{\displaystyle \partial
(q^i\circ\widetilde{\Phi})}{\displaystyle
\partial t^A}\Big\vert_{\mathbf{t}}\,.
$$\end{enumerate}
\end{corollary}

\subsection{Hamiltonian  formalism.}\label{Hamform}

In this subsection we will develop the Hamiltonian $k$-symplectic
formalism on Lie algebroids, in an analogous way that in the standard case

 Let $(E,\lcf\cdot,\cdot\rcf_E,\rho_E)$ be a Lie algebroid over a manifold
$Q$. For the Hamiltonian approach we consider the dual bundle,
$\tau^{\;*}:E^{*}\to Q$  of $E$.

\subsubsection{The manifold $\stackrel{k}{\oplus} E^{\;*}$.}

The
standard $k$-symplectic Hamiltonian formalism develops on the bundle
$(T^1_k)^*Q$ of $k^1$-covelocities of $Q$, that is, the Whitney sum
of $k$ copies of $T^*Q$. Passing to  Lie algebroids $E$ as a
substitute of the tangent bundle, its natural to think that the
analog of $(T^1_k)^*Q$ is the Whitney sum over $Q$ of $k$ copies of
the dual space $E^*$.

We denote by $$\stackrel{k}{\oplus} E^{*}=E^{*} \oplus
\stackrel{k}{\ldots} \oplus E^{*}\,,$$ the Whitney sum of $k$ copies
of the vector bundle $E^{*}$, the projection map
$$\widetilde{\tau}^{*}:\stackrel{k}{\oplus} E^{*}\to Q,$$ given by $
\widetilde{\tau}^{*}( {a}_{1_ {q}}^{\;*},\ldots, {a}_{k_ {q}}^{\;*})=  {q}$

If $(q^i,y_\alpha)$ are local coordinates on $(\tau^*)^{-1}(U)\subseteq E^*$, then the induced local coordinates $(q^i,y^A_\alpha)$ on $(\widetilde{\tau}^*)^{-1}(U)\subseteq \keh$ are given by
$$
q^i( {a}_{1_ {q}}^{\;*},\ldots, {a}_{k_ {q}}^{\;*})=q^i( {q})\,,\quad
y_\alpha^A( {a}_{1_ {q}}^{\;*},\ldots, {a}_{k_ {q}}^{\;*})=y_\alpha( {a}_{A_ {q}}^{\;*})\;.
$$

\subsubsection{The  Hamiltonian prolongation}\label{Hamprol}

For the description of the Hamiltonian $k$-symplectic  formalism on Lie algebroids we
consider the prolongation of a Lie algebroid $E$ over the fibration
$\widetilde{\tau}^*\colon\keh\to Q$, that is, (see Section \ref{prolong}),
\begin{equation}\label{teh}\teh=\{( {e}_{q}, {v}_{ \mathbf{{b}}^*_{q}})\in E\times T(\keh) /\;
\rho_E( {e}_{q})=T(\widetilde{\tau}^*)( {v}_{ \mathbf{b}^{\mathbf{*}}_{q}})\}\,.\end{equation}

Taking into account the description of the prolongation
$\mathcal{T}^EP$, (see for instance,
\cite{CLMMM-2006,LMM-2005,Mart-2001} or section \ref{prolong} in this paper), on the particular case
$P=E^*\oplus\stackrel{k}{\ldots}\oplus E^*$ we obtain:
\begin{enumerate}
\item $\teh\equiv E\times_{TQ}T(\keh)$ is a Lie algebroid over $\keh$ with projection
   $$\widetilde{\tau}_{\keh}\colon \teh\equiv
E\times_{TQ}T(\keh)\longrightarrow \keh$$ and Lie algebroid structure
$(\lcf\cdot,\cdot\rcf^{\widetilde{\tau}^*},\rho^{\widetilde{\tau}^*}\,)$
where the anchor map
$$\rho^{\widetilde{\tau}^*}=E\times_{TQ}T(\keh)\colon\teh\to T(\keh)$$ is
the canonical projection over the second factor.

We refer to this Lie algebroid as the {\it Hamiltonian prolongation}.

\item  If $(q^i,y_\alpha^A)$ denotes a local coordinate system of
   $\keh$ then the induced local coordinate system on
    $\teh\equiv E\times_{TQ}T(\keh)$ is given by
     $$(q^i,y_\alpha^A,z^\alpha,w_\alpha^A)_{1\leq i\leq n,\;1 \leq A\leq k,\; 1\leq \alpha\leq m}$$
where
\begin{equation}\label{local coord prol h}\begin{array}{lcllcl}
 q^i( {e}_{q}, {v}_{ \mathbf{{b}}_{q}^*}) &=& q^i(q)
  \;,\quad &
y_\alpha^A( {e}_{q}, {v}_{ {\mathbf{b}}_{q}^*})&=&y_\alpha^A( {\mathbf{b}}_{q}^*)\;,
\\\noalign{\medskip}
z^\alpha( {e}_{q}, {v}_{ {\mathbf{b}}_{q}^*})&=&y^\alpha( {e}_{q}) \;,\quad
& w_\alpha^A( {e}_{q}, {v}_{ \mathbf{{b}}_{q}^*})
&=& {v}_{ {\mathbf{b}}_{q}^*}(y_\alpha^A)\;. \\
\end{array}
\end{equation}

\item The set
$\{\mathcal{X}_\alpha,\mathcal{V}_A^\alpha\}$ given by
\begin{equation}\label{base h} \begin{array}{lccrcl}
\mathcal{X}_\alpha\colon &\keh&\to&\mathcal{T}^E(\keh) &\equiv &
E\times_{TQ}T(\keh)\\\noalign{\medskip}
 & {\mathbf{b}}^*_ {q}&\mapsto &
\mathcal{X}_\alpha( \mathbf{{b}}^*_ {q}) &=&
(e_\alpha( {q});\rho^i_\alpha( {q})\ds\frac{\partial
}{\partial q^i}\Big\vert_{ {\mathbf{b}}^*_ {q}})
\\\noalign{\bigskip}
\mathcal{V}_A^\alpha\,\colon &\keh&\to&\mathcal{T}^E(\keh)&\equiv &
E\times_{TQ}T(\keh)\\\noalign{\medskip}
 & {\mathbf{b}}^*_ {q}&\mapsto &
\mathcal{V}_A^\alpha( \mathbf{{b}}^*_ {q})
&=&( {0}_ {q};\ds\frac{\partial}{\partial
y_\alpha^A}\Big\vert_{ {\mathbf{b}}^*_ {q}})\,,
\end{array}\end{equation} is a local basis of $Sec(\teh)$ the set of sections of $\widetilde{\tau}_{\keh}$. (See (\ref{base k-prol})).

\item The anchor map $\rho^{\widetilde{\tau}^*}\colon\teh\to T(\keh)$ allows
 us to
associate a vector field to each section $\xi\colon\keh\to\teh$ of $\widetilde{\tau}_{\keh}$  .

Locally, if $\xi$ writes as follows:
$$\xi=\xi^\alpha\mathcal{X}_\alpha+\xi_\alpha^A\mathcal{V}_A^\alpha\in Sec(\teh)$$
then the associated vector field has the following local expression, see (\ref{rhoz}),
\begin{equation}\label{rholksim h}
\rho^{\widetilde{\tau}^*}(\xi)=\rho^i_\alpha \xi^\alpha\derpar{}{q^i}
+ \xi_\alpha^A\derpar{}{y_\alpha^A}\in \vf(\keh)\,.\end{equation}

\item The Lie bracket of two sections  of $\widetilde{\tau}_{\keh}$
is characterized by the following expressions (see (\ref{lie brack k-prol})):
\begin{equation}\label{lie brack te h}\begin{array}{lll}
\lcf\mathcal{X}_\alpha,\mathcal{X}_\beta\rcf^{\widetilde{\tau}^*}=
\mathcal{C}^\gamma_{\alpha\beta}\mathcal{X}_\gamma &
\lcf\mathcal{X}_\alpha,\mathcal{V}_B^\beta\rcf^{\widetilde{\tau}^*}=0
&
\lcf\mathcal{V}_A^\alpha,\mathcal{V}_B^\beta\rcf^{\widetilde{\tau}^*}=0\,,
\end{array}\end{equation}

\item If $\{\mathcal{X}^\alpha,
\mathcal{V}_\alpha^A\}$ is the dual basis of $\{\mathcal{X}_\alpha,
\mathcal{V}^\alpha_A\}$,  then the exterior differential is determined by, (see (\ref{dtp})),
\begin{equation}\label{difksim h}
\begin{array}{lcl}
d^{\teh}f&=&\rho^i_\alpha\derpar{f}{q^i}\mathcal{X}^\alpha +
\derpar{f}{y_\alpha^A}\mathcal{V}_\alpha^A\,,\quad \makebox{for all} \;
f\in \mathcal{C}^\infty(\keh)\\
d^{\teh}\mathcal{X}^\gamma &=&
-\ds\frac{1}{2}\mathcal{C}^\gamma_{\alpha\beta}\mathcal{X}^\alpha\wedge\mathcal{X}^\beta,\quad
d^{\teh}\mathcal{V}_\gamma^A =0\,.\end{array}\end{equation}
\end{enumerate}

\begin{remark}\label{remark equiv h}{\rm
In the particular case $E= TQ$ the manifold $\mathcal{T}^E(\keh)$
turns into $T(\tkqh)$.The proof is similar to remark \ref{remark equiv}.
 } \end{remark}

\subsubsection{The vector bundle $\mathcal{T}^E(\keh)\oplus\stackrel{k}{\ldots}\oplus\mathcal{T}^E(\keh)$.}

 In the standard $k$-symplectic Hamiltonian formalism one obtains the solutions of the Hamilton equations as integral sections of certain $k$-vector fields on $\tkqh$, that is, certain sections of
 \[
 \tau^k_{\tkqh}\colon T^1_k(\tkqh)\to \tkqh\,.
 \]

 Thinking on a Lie algebroid $E$ as a substitute of the tangent bundle, we know that $\teh$ plays the role of $T(\tkqh)$. Thus it is natural to choose the Whitney sum of $k$ copies of $\teh$, that is, the manifold
 \[
 (\mathcal{T}^E)^1_k(\keh)\colon =\teh\oplus\stackrel{k}{\ldots}\oplus\teh
 \] plays the role of
 \[T^1_k(\tkqh)=T(\tkqh)\oplus\stackrel{k}{\ldots}\oplus T(\tkqh)\,.
 \]

 We denote by $\widetilde{\tau}^k_{\keh}\colon  (\mathcal{T}^E)^1_k(\keh)\to\keh$ the canonical projection given by
 \[
 \widetilde{\tau}^k_{\keh}(Z^1_{ {\mathbf{b}}_{q}^*},\ldots, Z^k_{\mathbf{b}_q^*})= {\mathbf{b}}_{q}^*\,,
 \]
 where $Z^A_{ \mathbf{{b}}_{q}^*}=({ {a}_A}_{q},{ {v}_A}_{ \mathbf{{b}}_{q}^*})\in \mathcal{T}^E(\keh),\; A=1,\ldots, k$.

 Now, we consider a section $\mathbf{\xi}$ of $\widetilde{\tau}^k_{\keh}$. Next we will prove that there exist a $k$-vector field on $\keh$ associated to each section $\xi$.

 Notice that to give a section $$\mathbf{\xi}\colon \keh\to (\mathcal{T}^E)^1_k(\keh))=\teh\oplus\stackrel{k}{\ldots}\oplus \teh$$ of $\widetilde{\tau}^k_{\keh}$ is equivalent to give $k$ sections, $\xi_1,\ldots,\xi_k$ of the Hamiltonian prolongation $\teh$, obtained by projection $\mathbf{\xi}$ on each factor $\teh$.

 \begin{proposition}\label{vfsec}
 Let $\mathbf{\xi}=(\xi^1,\ldots, \xi^k)$ be a section of $\widetilde{\tau}^k_{\keh}$. Then $$(\rho^{\widetilde{\tau}^{\,*}}(\xi_1),\ldots,\rho^{\widetilde{\tau}^{\,*}}(\xi_k))\colon \keh\to
T^1_k(\keh)$$ is a $k$-vector field on $\keh$. Let us remember that the mapping
   $\rho^{\widetilde{\tau}^{\,*}}$ is the anchor map of the Lie algebroid
$\teh$.
 \end{proposition}
\proof Is a direct consequence of (\ref{rholksim h}) and the above remark.\qed

\subsubsection{Hamiltonian formalism}\label{Hamfor1}

Let $(E,\lcf\cdot,\cdot\rcf_E,\rho_E)$ be a Lie algebroid on a manifold $Q$
and $H:\stackrel{k}{\oplus} E^{\;*}\to \r$ be a Hamiltonian function.

In this subsection we will develop the $k$-symplectic Hamiltonian formalism on Lie algebroids. Morevover, we also describe the standard $k$-symplectic Hamiltonian formalism as a particular case of the formalism developed here.

 First, we define certain type of sections of the dual of the Hamiltonian prolongation $\teh$, which play the role of the Liouville forms in the standard case.

\paragraph{\bf \bf The Liouville sections} We are called {\it Liouville $1$-sections} to the sections of the bundle $(\teh)^*\to \keh$ defined as follow:
$$\begin{array}{rcc}
\Theta^A:\stackrel{k}{\oplus}E^{\,*} & \longrightarrow &
(\mathcal{T}^E(\stackrel{k}{\oplus}E^{\,*}))^{\;*}
\\\noalign{\medskip}
  \mathbf{{b}}_ {q}^{\;*} & \longmapsto & \Theta^A_{
  {\mathbf{b}}_ {q}^{\;*}}\end{array}\quad 1\leq A\leq k\,,
 $$
where  $\Theta^A_{  \mathbf{{b}}_ {q}^{\;*}}$ is the function given by
 \begin{equation}\label{thetaal}{\small\begin{array}{lrll}
 \Theta^A_{  \mathbf{{b}}_ {q}^{\;*}}: &
(\mathcal{T}^E(\stackrel{k}{\oplus}E^{\,*}) )_{
 \mathbf{{b}}_ {q}^{\;*}}& \longrightarrow & \r
\\\noalign{\medskip}
 & ( {  e}_{q}, {  v}_{  \mathbf{{b}}_ {q}^{\;*}}) & \longmapsto &
  \Theta^A_{  \mathbf{{b}}_ {q}^{\;*}}( {  e}_{q}, {  v}_{
   \mathbf{{b}}_ {q}^{\;*}})={  b}_{A_ {q}}^{\;*}({  e}_{ {q}})\;,
\end{array}}\end{equation}
for each $ {  e}_{q}\in
E,\, {\mathbf{b}}_{q}^*=({{  b}_1}_{q}^*,\ldots,{{  b}_k}_{q}^*)\in \keh$
and $ v_{  \mathbf{{b}}_ {q}^{\;*}}\in T_{
 \mathbf{{b}}_ {q}^{\;*}}(\keh)$.

Now we define the $2$-sections
$$\Omega^A:\stackrel{k}{\oplus}E^* \to
(\mathcal{T}^E(\stackrel{k}{\oplus}E^*))^{\;*}\wedge
(\mathcal{T}^E(\stackrel{k}{\oplus}E^*))^{\;*},\;1\leq A\leq k$$ by
$$
\Omega^A=- d^{\teh}\Theta^A\;,
$$where $d^{\teh}$ denotes the exterior differential on the Lie algebroid $\teh$, see  (\ref{difksim h}).

Next we will write the local expression of the sections
$\Theta^A$ and $\Omega^A$.

 Let $\{\mathcal{X}_\alpha,\;\mathcal{V}_B^\beta\}$ be a local basis of ${\rm Sec}(\teh)$ and
$\{\mathcal{X}^\alpha_A,\;\mathcal{V}^B_\beta\}$ its dual basis.
Then from  (\ref{base h})   we have
\begin{equation}\label{theta*ksim}
\Theta^A=\ds\sum_{\beta=1}^my^A_\beta\mathcal{X}^\beta\;,\quad 1\leq
A\leq k\,.
\end{equation}

Thus, from   (\ref{rholksim h}), (\ref{lie brack te h}),
(\ref{difksim h}) and (\ref{theta*ksim}) we obtain the local expression of $\Omega^A$, that is,
 \begin{equation}\label{omega A*ksim}
\Omega^A=\sum_{\beta}\mathcal{X}^\beta\wedge\mathcal{V}^A_\beta +
\ds\frac{1}{2} \sum_{\beta,\gamma,\delta}
\mathcal{C}^\delta_{\beta\gamma}y^A_\delta\mathcal{X}^\beta\wedge
\mathcal{X}^\gamma\;, \quad 1\leq A\leq k\;.
\end{equation}
\begin{remark}{\rm\
\begin{enumerate}
\item In the particular case $k=1$ the Liouville sections introduced here are the Liouville sections on Mechanics on Lie algebroids, see   E.
Martinez, \cite{CLMMM-2006,Mart-2001b}.

\item When $E= TQ$ and $\rho_{TQ}=id_{TQ}$,
then $$\Omega^A(X,Y)=\omega^A(X,Y)\,,\qquad 1\leq A\leq k$$
where $X,Y$ are   vector fiels on $\tkqh$ and $\omega^1,\ldots,$ $
\omega^k$ are the canonical $2$-forms of the standard $k$-symplectic Hamiltonian formalism.
\end{enumerate}
}\end{remark}

\paragraph{\bf The Hamilton equations.}

\begin{theorem}\label{alhamform}
Let $H:\stackrel{k}{\oplus}E^{\,*}\to \r$ be a Hamiltonian and
$$\mathbf{\xi}=(\xi_1,\ldots,\xi_k): \stackrel{k}{\oplus}E^* \to
(\mathcal{T}^E)^1_k(\stackrel{k}{\oplus}E^{\,*})\equiv \teh\oplus\stackrel{k}{\ldots}\oplus\teh$$ a section of
$\widetilde{\tau}^{\,k}_{\stackrel{k}{\oplus}E^{\,*}}$, or equivalently, $k$ sections of the Hamiltonian prolongation, $\teh$, such that
\begin{equation}\label{eq h}
 \ds\sum_{A=1}^k\imath_{\xi_A}\Omega^A=\d^{\teh} H\;.
\end{equation} If
$\psi:\rk\to \stackrel{k}{\oplus}E^{\,*}\,$  is an integral section of $\mathbf{\xi}$, then $\psi$ is a solution of the following system of partial differential equations
\begin{equation}\label{ham eq}
  \ds\frac{\partial \psi^i}{\partial t^A} = \rho^i_\alpha\ds\frac{\partial H}
  {\partial y^A_\alpha} \quad\makebox{and}\quad
  \ds\sum_{A=1}^k  \ds\frac{\partial \psi^A_\alpha}{\partial
  t^A} =-\left(\mathcal{C}^\delta_{\alpha\beta}\,\psi^B_\delta\,
  \derpar{H}{y^B_\beta}
+ \rho^i_\alpha \ds\frac{\partial H}{\partial q^i} \right) \,.
\end{equation}
\end{theorem}

\begin{remark}{\rm In the particular case $E=TQ$ and $\rho=id_{TQ}$, the equations (\ref{ham eq}) are the Hamilton field equations. Therefore these equations (\ref{ham eq}) are called {\it the Hamilton equations on Lie algebroids}.}\end{remark}

\proof Let $$\mathbf{\xi}=(\xi_1,\ldots,\xi_k):\stackrel{k}{\oplus}E^* \to
(\mathcal{T}^E)^1_k(\stackrel{k}{\oplus}E^{\,*})$$ be a section of
$\widetilde{\tau}^{\,k}_{\stackrel{k}{\oplus}E^{\,*}}$ such that
(\ref{eq h}) holds.

 Consider $\{\mathcal{X}_\alpha,\;\mathcal{V}_B^\beta\}$, a local basis of sections of
$\widetilde{\tau}_{\stackrel{k}{\oplus}E^{\,*}}:\mathcal{T}^E(\keh)\to
\keh$, then each $\,\xi_A,\; A=1,\ldots, k$ can be written as follow:
\begin{equation}\label{localxi}
\xi_A=\xi^\alpha_A\mathcal{X}_\alpha+(\xi_A)^B_\alpha
\mathcal{V}_B^\alpha\;,
\end{equation}

 {} From (\ref{difksim h}), (\ref{omega A*ksim}) and (\ref{localxi})
  we obtain that  (\ref{eq h}) is locally expressed as follow
\begin{equation}\label{echloc}\begin{array}{rcl}
 \xi^\alpha_B &=&
  \ds\frac{\partial H}{\partial y^B_\alpha}\, 
  \\\noalign{\medskip}
 \ds\sum_{A=1}^k(\xi_ A)^A_\alpha &=&
 -\left(\mathcal{C}^\delta_{\alpha\beta}\,y^C_\delta\,\derpar{H}{y^C_\beta}
+ \rho^i_\alpha \ds\frac{\partial H}{\partial q^i} \right)
\end{array}\end{equation}

Next, let $\psi:\rk\to
\stackrel{k}{\oplus}E^{\,*},\;\psi( {\mathbf{t}}) =
(\psi^i( \mathbf{{t}}),\psi^A_\alpha( {\mathbf{t}})) $ be an integral section of $\mathbf{\xi}$, that is, $\psi$ is an integral section of $(\rho^{\widetilde{\tau}^{\,*}}(\xi),\ldots,
\rho^{\widetilde{\tau}^{\,*}}(\xi))$ the $k$-vector field on $\keh$ associated to $\xi$. Thus the following expressions holds:
\begin{equation}\label{sint1ksim}
 \xi^\beta_A \rho^i_\beta = \ds\frac{\partial \psi^i}{\partial
 t^A}\;,\;
  (\xi_A)_\beta^B = \ds\frac{\partial \psi^B_\beta}{\partial
  t^A}\;.
\end{equation}

Finally, from (\ref{echloc}) and (\ref{sint1ksim})
 we deduce that  $\psi$ satisfies the following system of partial differential equations.
 $$\label{ham eq}
  \ds\frac{\partial \psi^i}{\partial t^A} = \ds\frac{\partial H}
  {\partial y^A_\alpha}\rho^i_\alpha \quad\makebox{and}\quad
  \ds\sum_{A=1}^k  \ds\frac{\partial \psi^A_\alpha}{\partial
  t^A} =-\left(\mathcal{C}^\delta_{\alpha\beta}\,\psi^A_\delta\,
  \derpar{H}{y^A_\beta}
+ \rho^i_\alpha \ds\frac{\partial H}{\partial q^i} \right) \,.
$$
\qed

\begin{remark}\label{5.44}{\rm \
\begin{enumerate}
\item When $E= TQ$ and
$\rho_{TQ}=id_{TQ}$ the equation (\ref{eq h}) is the geometric version of the Hamilton field equation in the standard $k$-symplectic formalism. This fact will be explain after.

\item In  the particular case $k=1$, this theorem summarized the Hamiltonian Mechanics on Lie algebroids, see section 3.2 in \cite{CLMMM-2006} or section
3.3 on \cite{LMM-2005}.\end{enumerate}}
\end{remark}

As a final remark in this subsection, it is interesting to point out that the standard Hamiltonian $k$-symplectic formalism is a particular case of the Hamiltonian formalism on Lie algebroids. In this case  $E= TQ$ and
$\rho_E=id_{TQ}$ as we have comment in the point $(i)$ of the  remark \ref{5.44}.  We have:

\begin{itemize}
\item The manifold $\stackrel{k}{\oplus}E^{\,*}$ is identified  with
$(T^1_k)^{\,*}Q$; $\mathcal{T}^{TQ}((T^1_k)^{\,*}Q)$ with
$T((T^1_k)^{\,*}Q)$; and $(\mathcal{T}^{TQ})^1_k((T^1_k)^{\,*}Q)$ with
$T^1_k((T^1_k)^{\,*}Q)$.

\item A section $$\xi:\stackrel{k}{\oplus}E^{\,*} \to
(\mathcal{T}^E)^1_k(\stackrel{k}{\oplus}E^{\,*})$$ corresponds to
a $k$-vector field  $\xi=(\xi_1,\ldots,\xi_k)$ on
$(T^1_k)^{\,*}Q$, that is, $\xi$ is a section of
$\tau^k_{(T^1_k)^{\,*}Q}:T^1_k((T^1_k)^{\,*}Q) \to (T^1_k)^{\,*}Q$.

\item Let $f$ be  a function defined on $(T^1_k)^{\,*}Q$ then
$$(d^{\teh}f)(Y)=df(Y)$$
where $df$ denotes the usual differential and $Y$ is a vector field on $(T^1_k)^{\,*}Q$.

\item It is satisfies that $$\Omega^A(X,Y)=\omega^A(X,Y)\quad (A=1,\ldots,k)$$
where $\omega^A, \; A=1,\ldots, k$ are the canonical $k$-symplectic $2$-forms on $(T^1_k)^*Q$.

\item Thus, in the standard Hamiltonian $k$-symplectic formalism the equation (\ref{eq h}) writes as follow:
$$\ds\sum_{A=1}^k\imath_{\xi_A}\omega^A=dH\,.$$
\end{itemize}

As consequence of the Theorem  \ref{alhamform} and the five above remarks, we reobtain the standard Hamiltonian $k$-symplectic formalism, which can be summarized in the following
\begin{corollary}Let $H:(T^1_k)^{\,*}Q \to \r$ be a Hamiltonian formalism and
$\mathbf{\xi}=(\xi_1,\ldots, \xi_k)$ be a $k$-vector field on
$(T^1_k)^{\,*}Q$ such that
$$\ds\sum_{A=1}^k\imath_{\xi_A}\omega^A=dH\,.$$

 If $\psi:\rk\to (T^1_k)^{\,*}Q ,\; \psi( \mathbf{{t}})=(\psi^i( {\mathbf{t}}),\psi^A_i( {\mathbf{t}}))$ is an integral section of $\mathbf{\xi}$, then is a solution to the Hamilton field equation in the standard $k$-symplectic formalism, that is, \begin{equation}\label{ecHksim} \ds\sum_{A=1}^k\frac{\ds \partial
\psi^A_i}{\ds \partial t^A}\Big\vert_{ \mathbf{t}}\,=
    \,- \frac{\ds \partial H}{\ds \partial q^i}\Big\vert_{
    \psi( \mathbf{t})}\,,\quad\frac{\ds
\partial \psi^i}{\ds
\partial t^A}\Big\vert_{  \mathbf{t}}\,=
    \, \frac{\ds \partial H}{\ds \partial p^A_i}\Big\vert_{ \psi( \mathbf{t})}\,
    \;,\quad i= 1\ldots, n\, .\end{equation}
\end{corollary}
\subsection{The Legendre transformation and the equivalence between the Lagrangian and Hamiltonian $k$-symplectic formalism on Lie algebroids}\protect\label{Sec 5.5.}

In this section we introduce the Legendre transformation on the $k$-symplectic framework on Lie algebroids and we  establish the equivalence between the Lagrangian and Hamiltonian formulation when we consider a hyperregular Lagrangian, This fact extends the analogous results of the standard case.

Let  $L:\stackrel{k}{\oplus}E\to\r$ be a Lagrangian function and
$\Theta_L^A\colon  \ke\to [\te]^*\;, \;(A=1,\ldots, k)$ be the Poincar\'{e}-Cartan
$1$-sections associated with  $L$, which was defined in
(\ref{theta al 2}).

\begin{definition}\label{legtrans} We introduce the {\it Legendre transformation associated with $L$} as the smooth map
$$\Le:\stackrel{k}{\oplus}E\to\stackrel{k}{\oplus} E^{\,*}$$ defined by
\[\mathfrak{Leg}( {b}_{1_ {q}},\ldots, {b}_{k_ {q}})=\Big([\Le( {b}_{1_ {q}},\ldots,
 {b}_{k_ {q}})]^1,\ldots, [\Le( {b}_{1_ {q}},\ldots,
 {b}_{k_ {q}})]^k\Big)\] where
\[[\Le( {b}_{1_ {q}},\ldots,
 {b}_{k_ {q}})]^A( {e}_{q} )=\ds\frac{d}{ds}L( {b}_{1_ {q}},\ldots,
{ {b}_A}_{ {q}}+s {e}_{q} ,\ldots,
 {b}_{k_ {q}})\Big\vert_{s=0}\,,\] being $ {e}_{q} \in
E_ {q}$.\end{definition}

 En other words, for each $A$ we can write
\begin{equation}\label{lea}[\Le( {b}_{1_ {q}},\ldots,
 {b}_{k_ {q}})]^A( {e}_{q} )=\Theta_L^A( {b}_{1_ {q}},\ldots,
 {b}_{k_ {q}})(Z)\,,\end{equation} where $Z$ is a point in the fiber of
$(\mathcal{T}^E(\stackrel{k}{\oplus}E))_{ \mathbf{{b}}_ {q}}$,
over the point
 $$ \mathbf{{b}}_ {q}=( {b}_{1_ {q}},\ldots,
 {b}_{k_ {q}})\in \stackrel{k}{\oplus}E$$ such that
$$\widetilde{\tau}_1(Z)= {e}_{q} $$ being
$$\widetilde{\tau}_1:\mathcal{T}^E(\stackrel{k}{\oplus}E)=E\times_{TQ}T(\ke)\to
E$$ is the projection over the first factor. Therefore $Z$ is of the form $Z=( {e}_{q}, {v}_{ {b}_ {q}})$.

The map $\Le$ is well-defined and its local expression is
\[\Le(q^i,y^\alpha_A)=(q^i,\ds\frac{\partial L}{\partial
y^\alpha_A})\,.\]

{}From this local expression it is easy to prove that the Lagrangian  $L$ is regular if an only if $\Le$ is a local diffeomorphism.

\begin{remark}{\rm When
$E=TQ$ the Legendre transformation defined here coincides with the Legendre transformation introduced by   G\"{u}nther in
\cite{Gu-1987}.  }\end{remark}

The Legendre transformation, $\mathfrak{Leg}$, induce a map
$$\mathcal{T}^E\Le:\mathcal{T}^E(\stackrel{k}{\oplus}E)\equiv E\times_{TQ}T(\ke)\to
\mathcal{T}^E(\stackrel{k}{\oplus}E^{\,*})\equiv
E\times_{TQ}T(\keh)$$ defined as follow
\[\mathcal{T}^E\Le(  {e}_{q},  {v}_{{  \mathbf{b}}_ {q}})=\Big(  {e}_{q},(\Le)_*({  \mathbf{b}}_ {q})(  {v}_{{  \mathbf{b}}_ {q}}) \Big)\,,\]  where
${  e}_ {q}\in
E_{q},\; {\mathbf{b}}_ {q}\in\stackrel{k}{\oplus}E$ y $(
 {e}_{q},  {v}_{{  \mathbf{b}}_ {q}}) \in
\mathcal{T}^E(\stackrel{k}{\oplus}E)\equiv E\times_{TQ}T(\ke)$.
Notice that the following diagram is commutative
\[\xymatrix{\stackrel{k}{\oplus}E\ar[rr]^-{\Le}\ar[rd]_-{\widetilde{\tau}}
 &&\stackrel{k}{\oplus}E^{\,*}\ar[dl]^-{\widetilde{\tau}^{\,*}}\\\ & Q&}
\] and thus $\mathcal{T}^E\Le$ is well-defined.

If we consider local coordinates on $\te$ (resp. $\teh$), see (\ref{local
coord prol}) and (\ref{local coord prol h}),  the local expression of $\mathcal{T}\Le$ is
\begin{equation}\label{teleg}\mathcal{T}^E\Le(q^i,y^\alpha_A,z^\alpha,
w^\beta_B)=(q^i,\ds\frac{\partial L}{\partial
y^\alpha_A},z^\alpha,z^\alpha\rho^i_\alpha\ds\frac{\partial^2
L}{\partial q^i\partial y^\gamma_C} + w^\beta_B\ds\frac{\partial^2
L}{\partial y^\gamma_C\partial y^\beta_B})\,.\end{equation}

\begin{theorem}\label{equivalencia forma al} The pair $(\mathcal{T}^E\Le, \Le)$ is a morphism between the Lie algebroids
$(\mathcal{T}^E(\stackrel{k}{\oplus}E),\rho^{\widetilde{\tau}},\lcf\cdot,\cdot\rcf^{\widetilde{\tau}})$
and
$(\mathcal{T}^E(\stackrel{k}{\oplus}E^*),\rho^{\widetilde{\tau}^{\,*}},\lcf\cdot,\cdot\rcf^{\widetilde{\tau}^{\,*}})$.
Moreover, if $\Theta_L^A$ and $\Omega_L^A$ (respectively, $\Theta^A$ and
$\Omega^A$) are the Poincar\'{e}-Cartan $1$-sections and $2$-sections associated with $L\colon \ke\to\r$ (respectively, the Liouville $1$-sections and $2$-sections on $\mathcal{T}^E(\stackrel{k}{\oplus}E^*)$),
then
\begin{equation}\label{equiv formas}
(\mathcal{T}^E\Le, \Le)^*\Theta^A=\Theta_L^A,\qquad
(\mathcal{T}^E\Le, \Le)^*\Omega^A=\Omega_L^A\,,\quad 1\leq A\leq
k\,.
\end{equation}
\end{theorem}

\proof Firstly we have to prove that $(\mathcal{T}^E\Le,
\Le)$ is a Lie algebroid morphism.
\[\xymatrix{\mathcal{T}^E(\stackrel{k}{\oplus}E)
\ar[rr]^-{\mathcal{T}^E\Le}\ar[d]_-{\widetilde{\tau}_{\stackrel{k}{\oplus}E}}
&&\mathcal{T}^E(\stackrel{k}{\oplus}E^{\,*})\ar[d]^-{\widetilde{\tau}_{\keh}}\\
\stackrel{k}{\oplus}E\ar[rr]^-{\Le} &&\stackrel{k}{\oplus}E^{\,*}
}\]

Suppose that  $(q^i)$ are local coordinates on $Q$, that $\{e_\alpha\}$ is a local basis of ${\rm Sec}(E)$ and denote by
$\{\mathcal{X}_\alpha,\mathcal{V}_\alpha^A\}$ (respectively,
$\{\mathcal{Y}_\alpha,\mathcal{U}^\alpha_A\}$) the corresponding local basis of sections of $\widetilde{\tau}_{\ke}\colon  \te\to \ke$
(respectively, $\widetilde{\tau}_{\keh}\colon  \teh\to \keh$).

Then, using (\ref{pullsec}), (\ref{difksim}) and (\ref{teleg}),
by a straightforward computation we deduce that
\begin{equation}\label{equivformas1}
(\mathcal{T}^E\Le,
\Le)^*(\mathcal{Y}^\alpha)=\mathcal{X}^\alpha\quad,\quad
(\mathcal{T}^E\Le,
\Le)^*(\mathcal{U}^A_\alpha)=d^{\te}\left(\derpar{L}{y^\alpha_A}\right)\,,\end{equation}
for each $\alpha=1,\ldots, m$ and $A=1,\ldots, k$ where
$\{\mathcal{X}^\alpha,\mathcal{V}^\alpha_ A\}$ and
$\{\mathcal{Y}^\alpha,\mathcal{U}_\alpha^A\}$ denotes the dual basis of $\{\mathcal{X}_\alpha,\mathcal{V}_\alpha^A\}$ and
$\{\mathcal{Y}_\alpha,\mathcal{U}^\alpha_A\}$  respectively.

Thus, taking into account this identities, from (\ref{difksim}) and
(\ref{difksim h}) we conclude that
$$\begin{array}{lcl}
(\mathcal{T}^E\Le, \Le)^*(d^{\teh} f) &=&
d^{\te}(f\circ\Le)\\\noalign{\medskip} (\mathcal{T}^E\Le,
\Le)^*(d^{\teh} \mathcal{Y}^\alpha) &=& d^{\te}((\mathcal{T}^E\Le,
\Le)^*\mathcal{Y}^\alpha)\\\noalign{\medskip} (\mathcal{T}^E\Le,
\Le)^*(d^{\teh} \mathcal{U}^A_\alpha) &=& d^{\te}((\mathcal{T}^E\Le,
\Le)^*\mathcal{U}^A_\alpha)\,,\end{array}$$ for all function $f\in
\mathcal{C}^\infty(\keh)$ and for all $\alpha$ and $A$.

Consequently, the pair
$(\mathcal{T}^E\Le, \Le)$ is a Lie algebroid morphism

Next we will check that
$(\mathcal{T}^E\Le,\Le)^*\Theta^A=\Theta_L^A$ holds.

{} From (\ref{pullsec}), (\ref{thetaal}) and (\ref{lea}) we obtain:
$$\begin{array}{lcl}
[(\mathcal{T}^E\Le,\Le)^*\Theta^A]_{ \mathbf{{b}}_{q}}( {e}_{q}, {v}_{ {\mathbf{b}}_{q}})&=&
\Theta^A_{\Le( {\mathbf{b}}_{q})}( {e}_{q},(\Le)_*( {\mathbf{b}}_{q})( {v}_{ {\mathbf{b}}_{q}}))\\\noalign{\medskip}&=&
[\Le( {\mathbf{b}}_{q})]^A( {e}_{q})=\Theta_L^A( {\mathbf{b}}_{q})( {e}_{q}, {v}_{ {\mathbf{b}}_{q}})\,.\end{array}
$$

Finally, since $(\mathcal{T}^E\Le, \Le)$ is a Lie algebroid morphism and taking into account the last identity we deduce that:
$$
(\mathcal{T}^E\Le,\Le)^*\Omega^A= \Omega_L^A\,. $$\qed

\begin{remark}{\rm In the particular case $k=1$ this theorem corresponds with the Theorem 3.12 of \cite{LMM-2005}.

In the case
$E=TQ$ and $\rho_{TQ}=id_{TQ}$ this theorem establishes the relation between the Lagrangian and Hamiltonian forms in the standard $k$-symplectic approach. }\end{remark}

Next, we will assume that $L$ is {\it hyperregular}, that is, $\Le$ is a global diffeomorphism. In this case we may consider the Hamiltonian function $H\colon \keh\to \r$ defined by
 $$H=E_L\circ (\Le)^{-1},$$ where $E_L$ es the Lagrangian energy associated with $L$ given by  (\ref{local ener}). Here
$(\Le)^{-1}$ is the inverse of the Legendre transformation
\[\xymatrix@=12mm{\keh\ar[r]^-{\Le^{-1}}\ar@{-->}[rd]_-{H}
&\ke\ar[d]^-{E_L}\\ & \r}\]
\begin{lemma}\label{difeo Leg}
If the Lagrangian $L$ is hyperregular then $\mathcal{T}^E\Le$
is a diffeomorphism
\end{lemma}
\proof The condition $L$ hyperregular means that $\Le$
is a global diffeomorphism, that is,  there exists its inverse map
$$\Le^{-1}\colon \stackrel{k}{\oplus}E^{\,*}\to\stackrel{k}{\oplus}E.$$

We define the inverse map to $\mathcal{T}^E\Le$ as the mapping
$$(\mathcal{T}^E\Le)^{-1}:\mathcal{T}^E(\stackrel{k}{\oplus}E^{\,*})\to
\mathcal{T}^E(\stackrel{k}{\oplus}E)$$ given by
\[(\mathcal{T}^E\Le)^{-1}(  {a}_ {q}, {v}_{ {\mathbf{b}}^{\,*}_ {q}})=\Big(  {a}_{q},(
\Le^{-1})_*( {\mathbf{b}}^{\,*}_ {q})( {v}_{ \mathbf{{b}}^{\,*}_ {q}} )\Big)\,,\]
where $  {a}_ {q}\in
E,\; {\mathbf{b}}^{\,*}_ {q}\in\stackrel{k}{\oplus}E^{\,*}$ and $(
 {a}_ {q}, {v}_{ {\mathbf{b}}^{\,*}_ {q}}) \in
\mathcal{T}^E(\stackrel{k}{\oplus}E^{\,*})\equiv
E\times_{TQ}T(\keh)$.

Therefore, $\mathcal{T}^E\Le$ is a diffeomorphism.\qed

The following theorem establishes the equivalence between the Lagrangian and Hamiltonian $k$-symplectic formulation on Lie algebroids.

\begin{theorem}\label{Equivalencia} Let $L$ be a hyperregular Lagrangian. There is a bijective correspondence between the set of maps  $\eta:\rk\to \stackrel{k}{\oplus}E$ such that $\eta$ is an integral section of a solution $\mathbf{\xi}_L$ of the geometric Euler-Lagrange equations (\ref{ec ge EL}) and the set of maps $\psi:\rk\to \stackrel{k}{\oplus}E^{\,*}$ which are integral sections of some solution $\mathbf{\xi}_H$ of the geometric Hamilton equations (\ref{eq h}).
\end{theorem}
\proof

The proof is similar  to the standard case, see \cite{tesissilvia}. An outline of the proof is the following:

Let $\mathbf{\xi}_L=(\xi_L^1,\ldots,
\xi_L^k):\stackrel{k}{\oplus}E\to(\mathcal{T}^E)^1_k(\stackrel{k}{\oplus}E)$
be a solution of the geometric Euler-Lagrange equations on Lie algebroids (\ref{ec ge EL}), then
$\mathbf{\xi}_H=(\xi_H^1,\ldots, \xi_H^k)$ where each
$$\xi_H^A=\mathcal{T}^E\Le\circ \xi_L^A\circ (\Le)^{-1}$$ is a solution of (\ref{eq h}).

Moreover, {\it if $\eta:\rk\to\stackrel{k}{\oplus} E$ is an integral section of $\mathbf{\xi}_L=(\xi_L^1,\ldots,\xi_L^k)$, then
$$\Le\circ\eta:\rk\to\stackrel{k}{\oplus} E^*$$ is an integral section of  $\mathbf{\xi}_H=(\xi_H^1,\ldots,\xi_H^k)$ being
$$\xi_H^A=\mathcal{T}^E\Le\circ \xi_L^A\circ (\Le)^{-1}.$$}

The converse is proved in a similar way.\qed

\begin{remark}{\rm If we rewrite the results of this subsection in the particular case  $k=1$ we obtain the equivalence between the Lagrangian an Hamiltonian Autonomous Mechanics on Lie algebroids, see for instance
\cite{CLMMM-2006}.

When $E= TQ$ and $\rho_{TQ}=id_{TQ}$, we obtain the equivalence between the Lagrangian and Hamiltonian formulation in the standard $k$-symplectic framework, see  \cite{tesissilvia}}
 \end{remark}

\section{Examples}

\paragraph{\bf Harmonic mappings} (\cite{CGR-2001,CR-2003,Van-2007,Wood-1994})
Here, we consider harmonic mappings $\phi:\r^2\to  G$ with
values in an arbitrary Lie group $ G$ with bi-invariant
metric $\langle\cdot,\cdot\rangle$. In the continuous case (see
\cite{Van-2007}), the harmonic mapping Lagrangian is given by
\begin{equation}\label{harm lag}
L(\phi,\phi_x,\phi_y)=\ds\frac{1}{2}\langle\phi^{-1}\phi_x,\phi^{-1}\phi_x\rangle +
 \ds\frac{1}{2}\langle\phi^{-1}\phi_y,\phi^{-1}\phi_y\rangle\,,
\end{equation}
 where $\langle\cdot,\cdot \rangle$ is the {\it Killing form} on
 $\mathfrak{g}$ and $\phi_x,\,\phi_y$ denotes the partial derivatives of $\phi$ respect to the local coordinates $(x,y)$ of $\r^2$ . The associated field equations are $\tau(\phi)=0$
 where $\tau(\phi)$ is the tension of $\phi$, defined as
 $$
 \tau(\phi)^i= h^{AB}\left(\frac{\partial^2 \phi^i}{\partial t^A\partial t^B} -
 \Gamma^C_{AB}\frac{\partial \phi^i}{\partial t^C} +
 \mathcal{C}^i_{jk}\frac{\partial\phi^j}{\partial
 t^A}\frac{\partial\phi^k}{\partial t^B}\right)\,,\quad
 i=1,\dots,\dim  G\,,
 $$where $h_{AB}$ are the components of a metric on $\r^2$, with
 Christoffel symbols $\Gamma^C_{AB}$ and $\mathcal{C}^i_{jk}$ are
 the Christoffel symbols of the bi-invariant metric on
 $G$. In our case, $h_{AB}$ is of course just the flat
 Euclidian metric.

 We will only treat the case of the harmonic maps that take values
 in $ G=SO(3)$, embedded in $\mathfrak{gl}(3)$, in which case the {\it
 Killing form} $\langle\cdot,\cdot \rangle$ is just the trace
 $$\langle\xi,\eta\rangle=-\,trace(\xi\eta)\,.
 $$

Let us observe that in this case, the Lagrangian (\ref{harm lag})
is represented as a function $L:TSO(3)\oplus TSO(3)\to \r$ defined
on $T^1_2(SO(3))$.

Taking into account that $T^1_2(SO(3))\cong
SO(3)\times\mathfrak{so}(3)\times\mathfrak{so}(3)$, we make the
identifications
$$T^1_2(SO(3))/SO(3)\cong \mathfrak{so}(3)\times\mathfrak{so}(3)$$
and we consider the  projection $l$ of $L$ to
$\mathfrak{so}(3)\times\mathfrak{so}(3)$ given by
$$
l(\xi_1,\xi_2)=-\frac{1}{2}trace(\xi_1^2) -
\frac{1}{2}trace(\xi_2^2)\,,\quad
\xi_1,\xi_2\in\mathfrak{so}(3)\,.$$

Let $\{E_1,E_2,E_3\}$ be a basis of $\mathfrak{so}(3)$, then
$\xi_i=y^\alpha_iE_\alpha,\; i=1,2$ and thus, $l$ is locally given
by $$l(y^\alpha_1,y^\alpha_2)= \ds\sum_{\alpha= 1}^3
\big((y^\alpha_1)^2 + (y^\alpha_2)^2\big)\,.$$

Since a Lie algebra is a example of a Lie algebroid we can apply the
theory developed in Section \ref{Lagform} and thus the
Euler-Lagrange (\ref{eq E-L ksim}) equations are given, in this case,  by
\[\begin{array}{rcl}\ds\frac{\partial y^\alpha_1}{\partial t^1} +\ds\frac{\partial y^\alpha_2}{\partial t^2} &=&
0\\\noalign{\medskip}
\ds\frac{\partial y^\alpha_A}{\partial t^B} - \frac{\partial y^\alpha_B}{\partial t^A} + \mathcal{C}^\alpha_{\beta\,\gamma}y^\beta_By^\gamma_A &=& 0\end{array}\quad (\alpha=1,2,3;\, A=1,2)\,\]

\paragraph{\bf \bf  Poisson sigma model.} Consider a Poisson manifold
$(Q,\Lambda)$. Then the cotangent bundle $T^*Q$ has a Lie algebroid structure, where the anchor is
$$\begin{array}{ccc}
\rho\colon T^*Q &\to & TQ\\\noalign{\medskip}
\beta &\mapsto &
\Lambda(\beta,\cdot)\end{array}$$ and the bracket is
\[[\alpha,\beta]=\imath_{\rho(\alpha)}d\beta -
\imath_{\rho(\beta)}d\alpha - d\Lambda(\alpha, \beta)\,.\]

In local coordinates, the bivector  $\Lambda$ has the local
expression
\[\Lambda = \ds\frac{1}{2}\Lambda^{ij}\ds\frac{\partial}{\partial
q^i}\wedge \ds\frac{\partial}{\partial q^j}\,.\]

We can consider the Lagrangian for the sigma Poisson model as a
function defined on $T^*Q\oplus T^*Q$. Thus if  $(q^i,p^i_1,p^i_2)$
denotes the local coordinates on $T^*Q\oplus T^*Q$, the local
expression of the Lagrangian is (see \cite{Mart})
\[L= -\,\frac{1}{2}\Lambda^{ij}p^1_ip^2_j\,.\]

A long but straightforward calculation shows that the Euler-Lagrange equation (\ref{eq E-L ksim}) are in this case

   $$\begin{array}{l}
   \ds\frac{1}{2}\Lambda^{ij}\left(\ds\frac{\partial p^2_i}{\partial t^1} - \ds\frac{\partial p^1_i}{\partial t^2}
   + \ds\frac{\partial \Lambda^{kl}}{\partial q^i} p^1_k p^2_l \right)=0\,,\\
  \ds\frac{\partial q^i}{\partial t^A} +\Lambda^{ij}p^A_j =0\;, \\
  \ds\frac{\partial p^2_i}{\partial t^1} - \ds\frac{\partial p^1_i}{\partial t^2}
   + \ds\frac{\partial \Lambda^{kl}}{\partial q^i} p^1_k p^2_l =0\;,
\end{array}$$
   In view of the morphism condition, we see that the first  equation vanishes. Thus the field equations are just
   $$\begin{array}{l}
  \ds\frac{\partial q^i}{\partial t^A} +\Lambda^{ij}p^A_j =0\;, \\
  \ds\frac{\partial p^2_i}{\partial t^1} - \ds\frac{\partial p^1_i}{\partial t^2}
   + \ds\frac{\partial \Lambda^{kl}}{\partial q^i} p^1_k p^2_l =0\;,
\end{array}$$ where a solution is a field $\phi:\r^2\to T^*Q\oplus
T^*Q$, locally given by $$\phi( {\bf t})=(q^i( {\bf t}),p^1_i( {\bf t}),p^2_i( {\bf t})).$$

Consider the $1$-forms on $\r^2$  given by $P_j=p^1_jdt^1 +
p^2_jdt^2 \;(j=1,\ldots, n)$, then the above equations can be
written  as
$$\begin{array}{l}
d\phi^j + \Lambda^{jk}P_k=0\\\noalign{\medskip} dP_j +
\ds\frac{1}{2} \Lambda_{,\,j}^{kl}P_k\wedge P_L=0 \,,\end{array}$$
that is the conventional form of the field equations for the
Poisson-sigma model \cite{Strobl-2004}

\begin{remark}
 Poisson sigma models   were originally introduced by
 Schaller, Strobl, \cite{SS-1994}, and Ikeda \cite{ikeda} so as to unify several
  two-dimensional models of gravity and to cast them into a
  common form with Yang-Mills theories.
\end{remark}

\paragraph{\bf \bf  Systems with symmetry.}

We consider a principal bundle $\pi: \bar{Q}\longrightarrow {Q}=\bar{Q}/G$. Let $A: T\bar{Q}\longrightarrow {\mathfrak g}$ be fixed  principal connection with curvature  $B: T\bar{Q}\oplus T\bar{Q}\longrightarrow {\mathfrak g}$.
The connection $A$ determines an isomorphism
between the vector bundles $T\bar{Q}/G\to Q$ and  $TQ\oplus \widetilde{\mathfrak g}\longrightarrow Q$ where $\widetilde{\mathfrak g}=(\bar{Q}\times {\mathfrak g})/G$ is the adjoint bundle (see \cite{CMR-2001}):
\[
[ {v}_{\bar{  q}}] \leftrightarrow T_{\bar{{  q}}}\pi( {v}_{\bar{{  q}}})\oplus [(\bar{{  q}}, A( {v}_{\bar{{  q}}}))]
\]
 where $ {v}_{\bar{{  q}}}\in T_{\bar{{  q}}}\bar{Q}$. The connection permits us to obtain a local basis of sections of ${\rm Sec} (T\bar{Q}/G)={\mathfrak X}(Q)\oplus {\rm Sec}( \widetilde{{\mathfrak g}})$ as follows. Let ${\mathfrak e}$ be the identity element of the Lie group $G$ and assume that there are local
coordinates $({q}^i)$, $1\leq i\leq \dim {Q}$ and that $\{\xi_a\}$ is a basis of ${\mathfrak g}$. The corresponding sections of the adjoint bundle  are the left invariant vector fields $\xi_a^L$:
\[
\xi_a^L(g)=T_{\mathfrak e} L_g(\xi_a)
\]
where $L_g: G\longrightarrow G$ is the left translation by $g\in G$.
If
\[
A\left(\frac{\partial}{\partial {q}^i}_{( {{q}}, {\mathfrak e})}\right)=A_i^a \xi_a
\]
then  corresponding horizontal lift on the trivialization $U\times G$ are the vector fields
\[
\left(\frac{\partial}{\partial {q}^i}\right)^h=\frac{\partial}{\partial {q}^i}-A_i^a\xi_a^L
\]

The set
\[
\left\{\left(\frac{\partial}{\partial {q}^i}\right)^h, \xi_a^L\right\}
\]
are by construction $G$-invariant and therefore, they constitute  a local basis of sections $\{ e_i, e_a\}$ of ${\rm Sec} (T\bar{Q}/G)={\mathfrak X}(Q)\oplus {\rm Sec}( \widetilde{\mathfrak g})$. Denote by $(q^i, y^i, y^a)$ the induced local coordinates of $T\bar{Q}/G$.
If ${\mathcal C}_{ab}^c$ are the structure constants of the Lie algebra
\[
B\left(\frac{\partial}{\partial {q}^i}_{( {{q}}, {\mathfrak e})},\frac{\partial}{\partial {q}^j}_{( {{q}}, {\mathfrak e})}\right)=
B_{ij}^a\xi_a
\]
where
\[
B_{ij}^c=\frac{\partial A^c_i}{\partial q^j}-\frac{\partial A^c_j}{\partial q^i}-{\mathcal C}^c_{ab}A^a_iA^b_j\; .
\]
then the structure functions of the  Lie algebroid $T\bar{Q}/G\rightarrow Q$ are determined by the following relations (see \cite{{LMM-2005}}):
\begin{eqnarray*}
\lcf e_i, e_j\rcf_{T\bar{Q}/G}&=& -B_{ij}^c e_c\\
\lcf e_i, e_a\rcf_{T\bar{Q}/G}&=&{\mathcal C}_{ab}^cA^b_i e_c\\
\lcf e_a, e_b\rcf_{T\bar{Q}/G}&=&{\mathcal C}_{ab}^c e_c\\
\rho_{T\bar{Q}/G}(e_i)&=&\frac{\partial}{\partial q^i}\\
 \rho_{T\bar{Q}/G}(e_a)&=&0\, .
 \end{eqnarray*}

 Now, consider a Lagrangian function $L: \stackrel{k}{\oplus}T\bar{Q}/G\longrightarrow \r$ then the Euler-lagrange field equations are:
 \begin{eqnarray*}
 \frac{d}{dt^A}
 \left(\frac{\partial L}{\partial y^i_A}\right)&=&\frac{\partial L}{\partial q^i}
 +B_{ij}^c y^j_C\frac{\partial L}{\partial y^c_C}-{\mathcal C}_{ab}^cA^b_i y^a_C\frac{\partial L}{\partial y^c_C}\\
 \frac{d }{d t^A}\left(\frac{\partial L}{\partial y^a_A}\right)&=&
 {\mathcal C}_{ab}^cA^b_i
   y^i_C\frac{\partial L}{\partial y^c_C}
 -{\mathcal C}_{ab}^cy^b_C\frac{\partial L}{\partial y^c_C}\\
 0&=&\frac{\partial y_A^i}{\partial t^B}-\frac{\partial y_B^i}{\partial t^A}\\
 0&=&\frac{\partial y_A^c}{\partial t^B}-\frac{\partial y_B^c}{\partial t^A}-B^c_{ij}y^i_By^j_A
 +{\mathcal C}_{ab}^cA^b_i
 y^i_By^a_A+{\mathcal C}_{ab}^cy^b_Ay^a_B
 \end{eqnarray*}

  In the case when $Q$ is a single point, that is $\bar{Q}=G$ then $T\bar{Q}/G={\mathfrak g}$ and then the Lagrangian is defined as a function
  $L: \stackrel{k}{\oplus}{\mathfrak g}\longrightarrow \r$ and then the previous equations are reduced now to
 \begin{eqnarray*}
  \frac{d }{d t^A}\left(\frac{\partial L}{\partial y^a_A}\right)&=&
  -c_{ab}^cy^b_C\frac{\partial L}{\partial y^c_C}\\
  0&=&\frac{\partial y_A^c}{\partial t^B}-\frac{\partial y_B^c}{\partial t^A}+{\mathcal C}_{ab}^cy^b_Ay^a_B
 \end{eqnarray*}
  which are a local expression of Euler-Poincar\'e equations, see for instance  \cite{CGR-2001} and \cite{Mart}.

\section*{Acknowledgments}
This work has been partially supported by MICIN (Spain) Grant MTM2008-00689, MTM 2007-62478, project ``Ingenio
Mathematica" (i-MATH) No. CSD 2006-00032 (Con\-so\-li\-der-In\-ge\-nio 2010)
and S-0505/ESP/0158 of the CAM.

\end{document}